\documentclass[aps,prx,twocolumn,superscriptaddress,longbibliography]{revtex4-1}
\usepackage{graphicx}
\usepackage[pdfpagemode=UseNone,pdfstartview=FitH,colorlinks=true,linkcolor=blue,citecolor=blue,urlcolor=blue]{hyperref}
\usepackage{placeins}
\usepackage[all]{hypcap}
\makeatletter
\let\saved@includegraphics\includegraphics
\AtBeginDocument{\let\includegraphics\saved@includegraphics}
\renewenvironment*{figure}{\@float{figure}}{\end@float}
\makeatother

\newcommand{\bra}[1]{\ensuremath{\left\langle#1\right|}}
\newcommand{\ket}[1]{\ensuremath{\left|#1\right\rangle}}

\begin{document}
\title{Scalable algorithm simplification using quantum AND logic}

\newcommand{\SIQSE}{\affiliation{1}{Shenzhen Institute for Quantum Science and Engineering, Southern University of Science and Technology, Shenzhen, Guangdong, China}}
\newcommand{\IQA}{\affiliation{2}{International Quantum Academy, Shenzhen, Guangdong, China}}
\newcommand{\GDKL}{\affiliation{3}{Guangdong Provincial Key Laboratory of Quantum Science and Engineering, Southern University of Science and Technology, Shenzhen, Guangdong, China}}
\newcommand{\DPHY}{\affiliation{4}{Department of Physics, Southern University of Science and Technology, Shenzhen, Guangdong, China}}
\newcommand{\ICT}{\affiliation{5}{Institute of Computing Technology, Chinese Academy of Sciences, Beijing, China}}
\newcommand{\UCAS}{\affiliation{6}{University of Chinese Academy of Sciences, Beijing, China}}

\author{Ji Chu}
\thanks{These authors have contributed equally to this work.}
\affiliation{\SIQSE}\affiliation{\IQA}\affiliation{\GDKL}
\author{Xiaoyu He}
\thanks{These authors have contributed equally to this work.}
\affiliation{\ICT}\affiliation{\UCAS}

\author{Yuxuan Zhou}
\affiliation{\SIQSE}\affiliation{\IQA}\affiliation{\GDKL}\affiliation{\DPHY}
\author{Jiahao Yuan}
\affiliation{\SIQSE}\affiliation{\IQA}\affiliation{\GDKL}\affiliation{\DPHY}
\author{Libo Zhang}
\affiliation{\SIQSE}\affiliation{\IQA}\affiliation{\GDKL}

\author{Qihao Guo}
\affiliation{\SIQSE}\affiliation{\IQA}\affiliation{\GDKL}
\author{Yongju Hai}
\affiliation{\SIQSE}\affiliation{\IQA}\affiliation{\GDKL}
\author{Zhikun Han}
\affiliation{\SIQSE}\affiliation{\IQA}\affiliation{\GDKL}
\author{Chang-Kang Hu}
\affiliation{\SIQSE}\affiliation{\IQA}\affiliation{\GDKL}
\author{Wenhui Huang}
\affiliation{\SIQSE}\affiliation{\IQA}\affiliation{\GDKL}\affiliation{\DPHY}
\author{Hao Jia}
\affiliation{\SIQSE}\affiliation{\IQA}\affiliation{\GDKL}
\author{Dawei Jiao}
\affiliation{\SIQSE}\affiliation{\IQA}\affiliation{\GDKL}
\author{Yang Liu}
\affiliation{\SIQSE}\affiliation{\IQA}\affiliation{\GDKL}
\author{Zhongchu Ni}
\affiliation{\SIQSE}\affiliation{\IQA}\affiliation{\GDKL}\affiliation{\DPHY}
\author{Xianchuang Pan}
\affiliation{\SIQSE}\affiliation{\IQA}\affiliation{\GDKL}
\author{Jiawei Qiu}
\affiliation{\SIQSE}\affiliation{\IQA}\affiliation{\GDKL}\affiliation{\DPHY}
\author{Weiwei Wei}
\affiliation{\SIQSE}\affiliation{\IQA}\affiliation{\GDKL}
\author{Zusheng Yang}
\affiliation{\SIQSE}\affiliation{\IQA}\affiliation{\GDKL}
\author{Jiajian Zhang}
\affiliation{\SIQSE}\affiliation{\IQA}\affiliation{\GDKL}\affiliation{\DPHY}
\author{Zhida Zhang}
\affiliation{\SIQSE}\affiliation{\IQA}\affiliation{\GDKL}\affiliation{\DPHY}
\author{Wanjing Zou}
\affiliation{\SIQSE}\affiliation{\IQA}\affiliation{\GDKL}

\author{Yuanzhen Chen}
\affiliation{\SIQSE}\affiliation{\IQA}\affiliation{\GDKL}\affiliation{\DPHY}
\author{Xiaowei Deng}
\affiliation{\SIQSE}\affiliation{\IQA}\affiliation{\GDKL}
\author{Xiuhao Deng}
\affiliation{\SIQSE}\affiliation{\IQA}\affiliation{\GDKL}
\author{Ling Hu}
\affiliation{\SIQSE}\affiliation{\IQA}\affiliation{\GDKL}
\author{Jian Li}
\affiliation{\SIQSE}\affiliation{\IQA}\affiliation{\GDKL}
\author{Dian Tan}
\affiliation{\SIQSE}\affiliation{\IQA}\affiliation{\GDKL}
\author{Yuan Xu}
\affiliation{\SIQSE}\affiliation{\IQA}\affiliation{\GDKL}
\author{Tongxing Yan}
\affiliation{\SIQSE}\affiliation{\IQA}\affiliation{\GDKL}

\author{Xiaoming Sun}
\email{sunxiaoming@ict.ac.cn}
\affiliation{\ICT}\affiliation{\UCAS}
\author{Fei Yan}
\email{yanf7@sustech.edu.cn}
\affiliation{\SIQSE}\affiliation{\IQA}\affiliation{\GDKL}
\author{Dapeng Yu}
\affiliation{\SIQSE}\affiliation{\IQA}\affiliation{\GDKL}\affiliation{\DPHY}

\date{}

\begin{abstract}
	Implementing quantum algorithms on realistic hardware requires translating high-level global operations into sequences of native elementary gates, a process known as quantum compiling. 
	Physical limitations, such as constraints in connectivity and gate alphabets, often result in unacceptable implementation costs.
	To enable successful near-term applications, it is crucial to optimize compilation by exploiting the potential capabilities of existing hardware.
	Here, we implement a resource-efficient construction for a quantum version of AND logic that can reduce the cost, enabling the execution of key quantum circuits. 
	On a high-scalability superconducting quantum processor, we demonstrate low-depth synthesis of high-fidelity generalized Toffoli gates with up to 8 qubits and Grover's search algorithm in a search space of up to 64 entries; both are the largest such implementations in scale to date.
	Our experimental demonstration illustrates a scalable implementation of simplifying quantum algorithms, paving the way for larger, more meaningful quantum applications on noisy devices.
\end{abstract}

\maketitle

Quantum algorithms are predicted to provide a computational speed-up over their classical counterparts.
To be implemented, these algorithms need to be compiled on specific quantum hardware to decompose global operations into the naturally available elementary gates.
Given the stringent resource constraints offered by the noisy intermediate-scale quantum (NISQ) technology foreseeable in the next 5--10 years \cite{preskill2018quantum}, it is essential to optimize the use of every qubit and every gate cycle to enable successful near-term applications \cite{chong2017programming}.
One effective strategy is to fully explore the hardware capabilities and diversify the available gate alphabets to optimize compilation \cite{foxen2020demonstrating, abrams2020implementation, gu2021fast}.

Several global or multi-qubit operations are textbook circuit components essential for building quantum algorithms \cite{nielsen2002quantum}. 
The best-known examples are the quantum arithmetic circuits used in Shor's factoring algorithm \cite{shor1994algorithms} and the multiply controlled gates used in Grover's search algorithm \cite{grover1997quantum}. The latter are nontrivial multi-qubit quantum logics that perform unitary operations on target qubits conditioned on the states of all the control qubits. Relevant applications include quantum error correction \cite{cory1998experimental, dennis2001toward, inada2021measurement}, quantum simulation \cite{childs2018toward}, and quantum machine learning \cite{tacchino2019artificial}.
One brute-force approach for an extensible implementation of these large operations is to decompose them into a finite set of universal gates.
For example, the generalized Toffoli gate, i.e., the $n$-qubit controlled-NOT (CNOT) gate, can be constructed using quadratically many ($\mathcal{O}(n^2)$) two-qubit CNOT gates plus single-qubit gates on a qubit array with all-to-all connections \cite{barenco1995elementary} and even more gates on devices with nearest-neighbor couplings \cite{mandviwalla2018implementing}. 
A more efficient approach is to concatenate together small Toffoli gates, assisted by ancilla qubits \cite{nielsen2002quantum, maslov2016advantages}. 
Leaving aside the extra resources needed, it is challenging to achieve high-quality small Toffoli gates. Apart from brute-force decomposition, small Toffoli gates may be obtained via one-step manipulations \cite{reed2012realization, song2017continuous, li2019realisation, levine2019parallela, roy2020programmable, hendrickx2021fourqubit, kim2021highfidelity} or by leveraging either, again, ancilla qubits \cite{figgatt2017complete, gidney2021cccz} or ancilla levels \cite{lanyon2009simplifying, mariantoni2011implementing, fedorov2012implementation, hill2021realization, galda2021implementing}.
Despite successful demonstrations of single small Toffoli gates in various systems, a scalable synthesis has never been experimentally realized because of the prohibitive implementation cost.
A scheme that is, at the same time, hardware-efficient, low-depth, easy to control, and compatible with state-of-the-art hardware \cite{arute2019quantum, mooney2021whole, wu2021strong} is yet to be realized.

In this study, we introduce a quantum version of the AND (QuAND) gate, a novel gate type that, as inspired by \cite{fedorov2012implementation}, utilizes an ancilla level for temporary information storage only.
The QuAND gate enables a scaling advantage in the circuit depth when synthesizing arithmetic circuits and multiply controlled gates. 
We experimentally implement QuAND gates on a superconducting quantum processor featuring simplified wiring and low crosstalk and demonstrate a linear-depth synthesis of generalized Toffoli gates with up to 8 qubits, i.e., a total of 7 control qubits.
To the best of our knowledge, our demonstration is the largest in size and the highest in performance to date (truth-table fidelity: 89.1\%, 53.2\%, and 39.1\% for $n=4,6$, and $8$, respectively). 
Using these gates, we perform Grover's search algorithm with multiple amplification cycles and achieve significant success probabilities (46.8\% and 3.9\% for searches of 16 and 64, respectively), demonstrating the feasibility of our method for scaled applications.
Note that alternative efficient compilation schemes have been proposed in recent theoretical studies \cite{gokhale2019asymptotic, inada2021measurement}; however, these schemes generally require the manipulation of a multi-level system with high degrees of freedom, adding considerable operational complexity.


\begin{figure}[htbp]
	\centering
	\includegraphics[width=85mm]{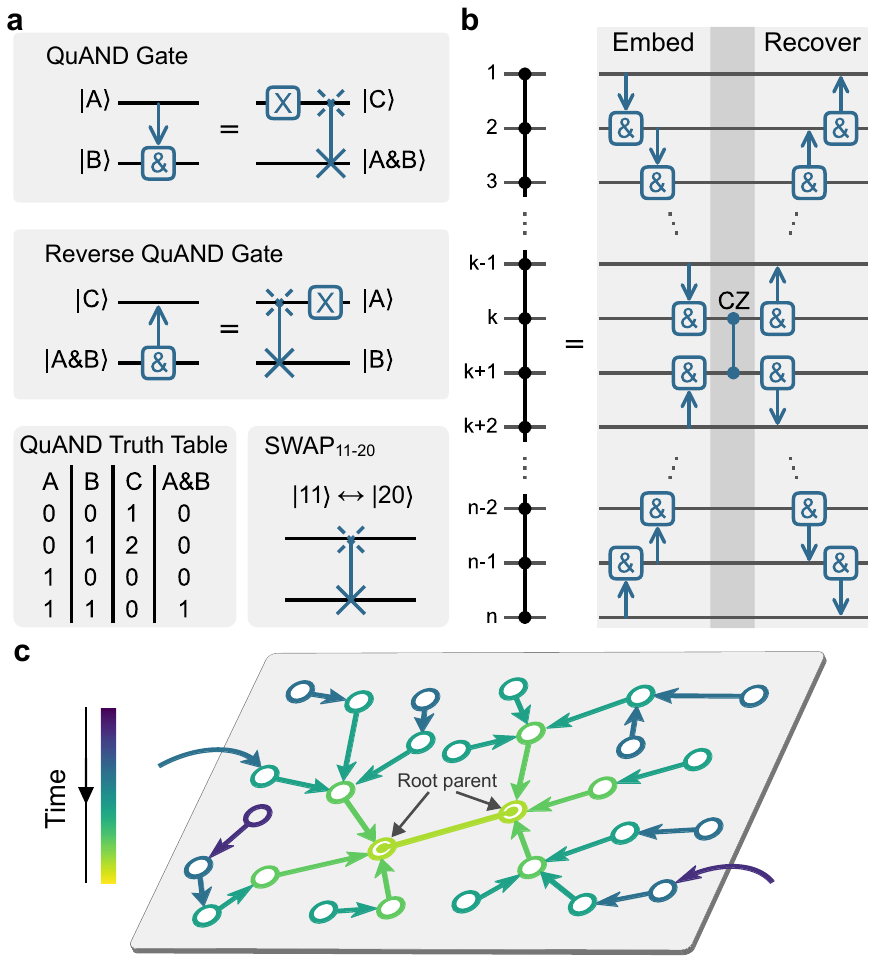}
	\caption{
		\textbf{Simplifying compilation using the quantum version of the AND (QuAND) gate.} 
		\textbf{a,} Circuit notation, truth table, and decomposition of the QuAND gate and its reversal.
		The AND-value qubit, indicated by an $\&$, is referred to as the ``parent'' and the other qubit is referred to as the ``child.'' A QuAND gate is indicated by an arrow pointing from the child to the parent, with an arrow in the opposite direction indicating a reverse QuAND gate. Both can be synthesized with a single-qubit $X$ gate and a SWAP operation between \ket{11} and \ket{20}, which is indicated by a double-cross sign with a dashed cross on the child qubit.
		\textbf{b,} Circuit decomposition of an $n$-qubit controlled-$Z$ (CZ) gate on a one-dimensional qubit chain using a sequence of QuAND gates, a CZ gate, and a sequence of reverse QuAND gates, shown here with time progressing from left to right.
		During embedding, the sequentially applied QuAND gates register the AND results of all the qubits from the upper and lower halves of the chain onto the two root parents, Q$_{k}$ and Q$_{k+1}$, respectively. The embedded information is later released via the reverse QuAND gates to recover the original binary encoding.
		The CZ gate is only effective when all qubits are in state \ket{1}. 
		\textbf{c,} Sketch showing a quantum processor with qubits connected in an arbitrary topology. A branching tree is enacted for implementing the QuAND gate sequence (arrows) with time progressing from dark blue to light green. The CZ gate is performed between the two root parents.
		The QuAND gate could also be performed across multiple processors (arrows pointing from outside) to efficiently implement global operations on a larger quantum network. 		
	}
	\label{fig:quand}
\end{figure}

The logic AND operation is a basic ingredient for designing both classical and quantum algorithms.
Unfortunately, it cannot be directly implemented on qubits because of the reversibility of quantum operations.
One workaround is to extract it from a Toffoli gate at the cost of an extra qubit \cite{nielsen2002quantum}; such overhead hinders scaled implementation on realistic hardware.
Here, we propose a resource-efficient QuAND gate scheme (Fig.~\ref{fig:quand}a) in which one of the two outputs registers the AND result of the inputs, i.e., $\ket{\rm A\&B}$, and the other output \ket{\rm C} spans three different states, in our case, \ket{1}, \ket{2}, and \ket{0} for the input states \ket{00}, \ket{01}, and \ket{10}, respectively.
The use of the ancilla level \ket{2} preserves the reversibility; the reverse QuAND gate simply switches the inputs and outputs.
We refer to the AND-value qubit as the ``parent'' and the other qubit as the ``child''.  
The circuit notation, truth table, and decomposition schemes of the QuAND gate and its reversal are illustrated in Fig.~\ref{fig:quand}a.
Here, we decompose the QuAND gate (or its reversal) into a single-qubit $X$ gate in front of (or after) a SWAP operation between \ket{11} and \ket{20} (denoted SWAP$_{11-20}$), which is naturally available on our hardware, as shown later.

One direct application of a QuAND gate is to simplify the compilation of large gate operations, in particular, multiply controlled gates, which are experimentally challenging to realize and the focus of this study.
Figure \ref{fig:quand}b shows the circuit decomposition for an $n$-qubit CZ gate on a one-dimensional qubit chain divided into three stages: embedding, the controlled-unitary operation, and recovery. 
Let $\ket{s}=\ket{s_1s_2...s_n}$ ($s_i=0,1$) denote a basis state at the input.
During embedding, we apply QuAND gates sequentially to the chain from both ends inward. 
At the end of the QuAND sequence, the two root parents in the middle, Q$_{k}$ and Q$_{k+1}$, temporarily register the AND result of all the qubits from the upper and lower halves of the chain, respectively. That is, $\ket{s_k'} = |\bigwedge_{i=1}^{k} s_i\rangle$ and $\ket{s_{k+1}'} = |\bigwedge_{i=k+1}^{n} s_i\rangle$, where $\bigwedge$ is the notation for the global AND operation.
Therefore, the subsequent CZ gate, which flips the wavefunction sign when $\ket{s_k' s_{k+1}'} = \ket{11}$, is effectively a phase flip conditioned on all qubits, $(-1)^{\bigwedge_{i=1}^{n} s_i}$.
The recovery sequence then transfers the state back to the original binary encoding and completes an $n$-qubit CZ gate.
There are a total of $2n-3$ two-qubit gates in this sequence.
The linear ($\mathcal{O}(n)$) circuit depth (the number of two-qubit gate cycles) as a result of using QuAND gates manifests a scaling advantage over the quadratic depth when using only CNOT gates \cite{barenco1995elementary}. 
Note that the ancilla levels are used here only for the temporary storage of the state information and that only a state-transfer operation is needed, which is in contrast to schemes that require more complex, hard-to-engineer operations with ancilla levels \cite{gokhale2019asymptotic, galda2021implementing, hill2021realization}. 
Moreover, other types of multiply controlled gates such as the generalized Fredkin (controlled-SWAP) gate can be similarly synthesized.
In fact, any classical circuit, such as Boolean logic and arithmetic circuits, can be constructed efficiently using QuAND and single-qubit gates because the classical NAND gate is universal. Examples of quantum adder circuits are shown in \cite{supplement}.

An even more impressive scaling advantage can be achieved on qubit arrays with higher connectivity.
To see this, it is helpful to first identify the key idea of our proposal. That is, to enact a branching tree graph on an arbitrarily connected qubit array and apply QuAND gates sequentially to register the AND results of neighboring qubits onto the parents layer-by-layer from the leaves up to the root, as illustrated in Fig.~\ref{fig:quand}c.
Ignoring the constant, the optimal circuit depth is then equivalent to the depth of the tree.
For example, the circuit depth can be reduced to $\mathcal{O}(\sqrt{n})$ on a two-dimensional square array and to $\mathcal{O}(\log_2n)$ on a binary tree \cite{supplement}; such polynomial or exponential speed-up in compiling global operations can constitute a huge boost for relevant quantum applications.
In addition, because this scheme only requires that qubits be connected, it is well suited for a distributed quantum network where only sparse connections are likely to be available.

\begin{figure}[htbp]
	\centering
	\includegraphics[width=85mm]{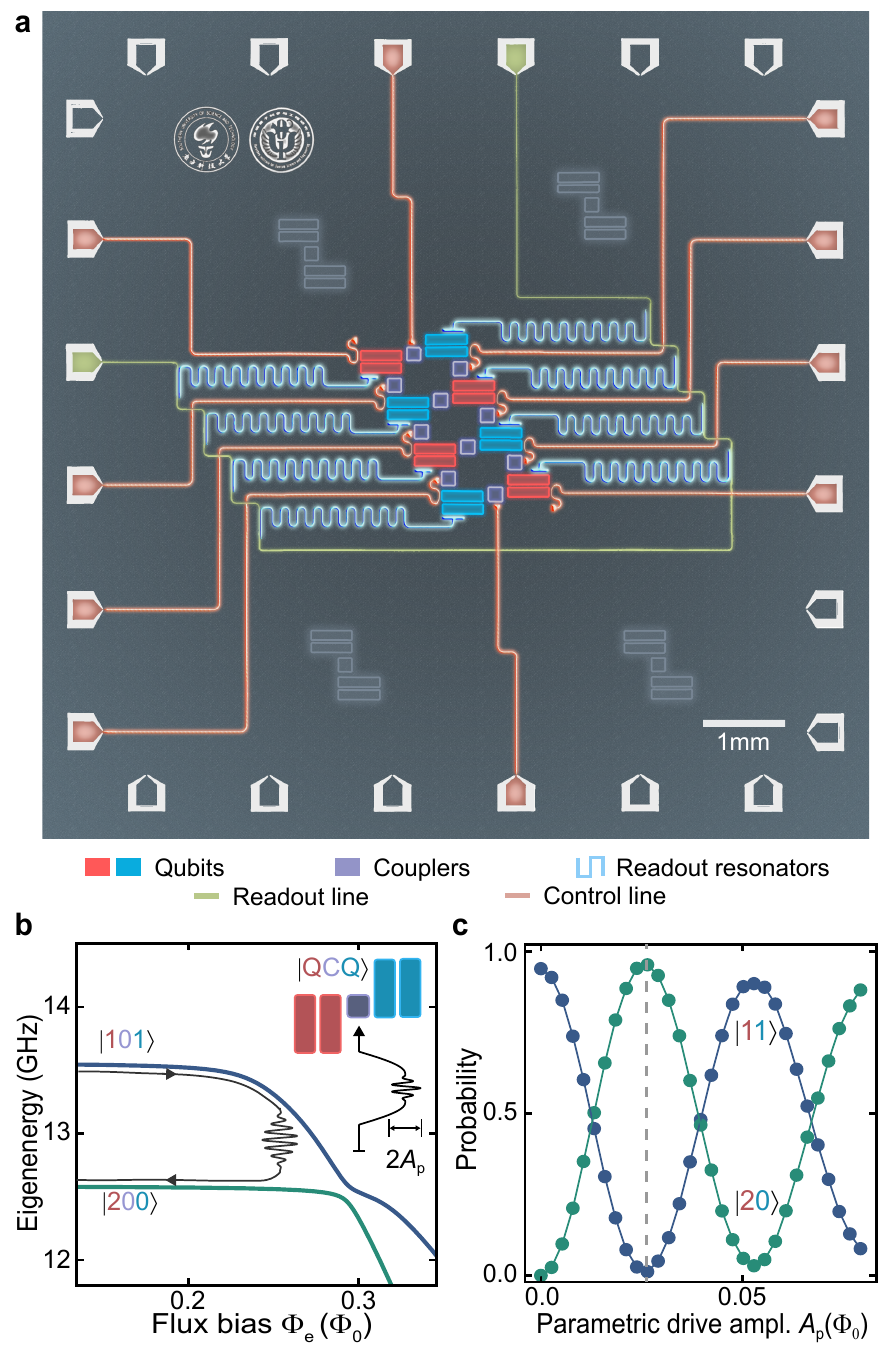}
	\caption{
		\textbf{Implementing the QuAND gate on a high-scalability superconducting quantum processor.}
		\textbf{a,} False-color micrograph of the device.
		Here, red and blue indicate the lower and higher fixed-frequency transmon qubits, respectively. 
		\textbf{b,} Eigenenergies of the states \ket{101} and \ket{200} (tri-mode notation) in a qubit--coupler--qubit subsystem versus the coupler-flux bias, $\Phi_{\rm e}$.
		The thin black line with the embedded arrows indicates the state trajectory for the SWAP$_{11-20}$ pulse sent to the coupler (inset), where $A_{\rm p}$ is the amplitude of the parametric drive on the flux pulse plateau. 
		\textbf{c,} Measured final state probabilities in states \ket{11} and \ket{20} after the SWAP$_{11-20}$ pulse versus the parametric drive amplitude.
		The dashed line indicates a full SWAP$_{11-20}$ operation. 
	}
	\label{fig:swap}
\end{figure}


Our experimental device (Fig.~\ref{fig:swap}a), tested inside a dilution refrigerator at a base temperature of 10~mK, consists of 8 fixed-frequency transmon qubits \cite{koch2007charge}, known for long coherence and simplified control, arranged in a ladder array and interconnected via 10 frequency-tunable couplers.
The two couplers in the middle have no control lines, resulting in the qubit array having a ring topology.
Each qubit has a dedicated readout resonator, and all the resonators share a common feed line enabling a multiplexed dispersive readout.
The qubit frequencies are arranged alternatively between a red band (6.2--6.5~GHz) and a blue band (7.0--7.3~GHz) along the ring; such frequency planning helps suppress the microwave crosstalk.
The qubits are strongly coupled (with an interaction strength of $g/2\pi \approx 100$~MHz) to their adjacent couplers, which are tunable via their flux biases $\Phi_{\rm e}$.
The couplers are designed to turn off the inter-qubit coupling via multi-path interference \cite{yan2018tunable} near their maximum frequencies (8.0--8.4~GHz) at $\Phi_{\rm e}=0$, which resolves the frequency-crowding problem and reduces the nearest-neighbor $ZZ$ crosstalk down to about 50~kHz.
In addition, the use of tunable couplers enables fast two-qubit gates between the fixed-frequency qubits, e.g., the adiabatic CZ gate \cite{collodo2020implementation, xu2020high}.
We use a shared control line to deliver the diplexed signals for both the qubit (4--8~GHz) and coupler (DC-1~GHz) control;
these signals are synthesized at room temperature and transmitted to the device inside the refrigerator. 
This design substantially simplifies the wiring effort both on the chip and inside the refrigerator, promising higher scalability.
See \cite{supplement} for details concerning the device and experimental setup.

\begin{figure*}[htbp]
	\centering
	\includegraphics[width=180mm]{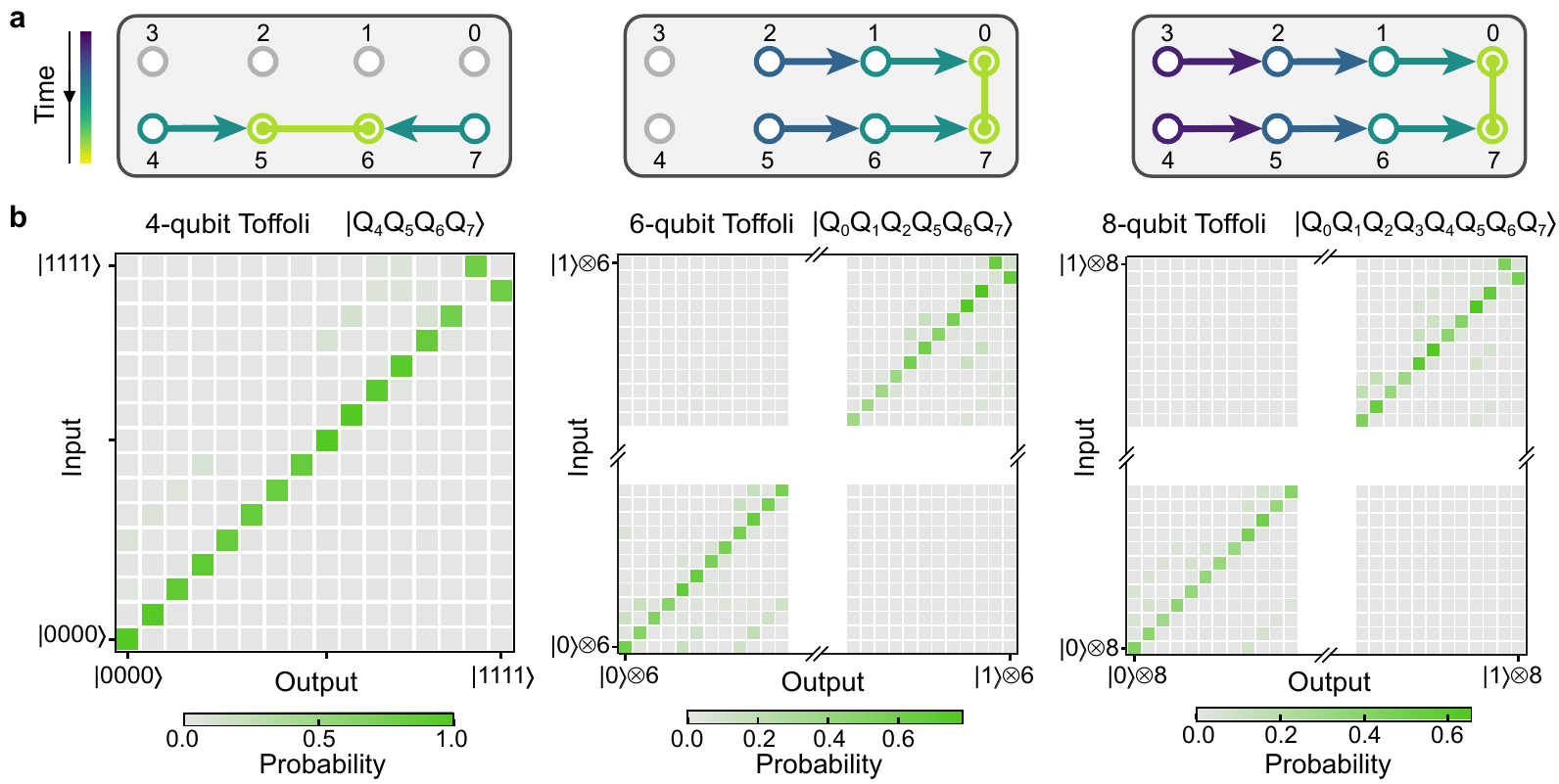}
	\caption{
		\textbf{Low-depth synthesis of generalized Toffoli gates.} 
		\textbf{a,} Schematic diagram showing the compiled sequence for implementing the 4-qubit (left), 6-qubit (center), and 8-qubit (right) CZ gate circuit described in Fig.~\ref{fig:quand}b on the 8-qubit processor with qubits indexed from 0 to 7.
		The arrows denote the QuAND gate sequence with time progressing from dark blue to light green. We have omitted the reverse QuAND sequence.
		\textbf{b,} Measured truth tables of the corresponding generalized Toffoli gate. In these examples, the controlled-NOT operation is performed on $\rm{Q}_7$ in all cases.
	}
	\label{fig:toffoli}
\end{figure*}

The QuAND and SWAP$_{11-20}$ gates on our device were implemented using coupler-assisted level transitions.
According to the tri-mode (\ket{{\rm qubit,coupler,qubit}}) notation, the SWAP$_{11-20}$ gate is a full swap operation between \ket{101} and \ket{200}, which is realized by a flux pulse sent to the coupler.
To activate such a transition, we applied a flux pulse, consisting of an adiabatic rise and fall (40~ns each) separated by a sinusoidal pulse (30~ns), to the coupler, as illustrated in Fig.~\ref{fig:swap}b.
Under this pulse, as shown by the thin black line with embedded arrows, the system state first follows an adiabatic excursion on state \ket{101} from the idling bias $\Phi_{\rm e} = 0~\Phi_0$ to $\Phi_{\rm e} = 0.26~\Phi_0$, then transits to \ket{200} via a parametric drive resonant with the instantaneous frequency gap between \ket{101} and \ket{200}, and eventually adiabatically returns to the idling bias. 
There are two major concerns when choosing the transition bias. First, the flux-induced \ket{101}$\leftrightarrow$\ket{200} transition is inhibited at $\Phi_{\rm e} = 0$ but is significantly enhanced at a sufficiently large bias as a result of wavefunction hybridization, as is evident by the strong bending of the energy levels \cite{chu2021coupler}.
Second, a proper bias is critical to avoid spurious transitions \cite{supplement}.

In the experiment, we calibrated the SWAP$_{11-20}$ gate by optimizing both the frequency and the amplitude of the parametric pulse. An example of the continuous swapping between \ket{11} and \ket{20} as a function of the pulse amplitude $A_{\rm p}$ is shown in Fig.~\ref{fig:swap}c.
The average observed transition error of 2.7\% is primarily caused by energy relaxation during the pulse.
All data presented here were corrected to account for the state preparation and measurement error. 
See \cite{supplement} for details concerning the gate scheme, pulse calibration, and data processing.


Using calibrated QuAND gates, we demonstrate the low-depth synthesis of a generalized Toffoli gate, which is equivalent to the $n$-qubit CZ circuit described in Fig.~\ref{fig:quand}b with two additional single-qubit gates. 
Figure \ref{fig:toffoli}a illustrates how we compile, on the 8-qubit ring, an $n$-qubit CZ gate with incremental size ($n=4, 6$, and $8$) in linear time steps.
We characterize these large gates by measuring their truth tables $U_{\rm exp}$, i.e., the output state probability distribution for each of the $2^n$ input states, which are shown in Fig.~\ref{fig:toffoli}b.
The truth-table fidelities $\mathcal{F}_{\rm tt} = \frac{1}{2^{n}} \rm{Tr}(U_{\rm exp} U_{ideal})$ are 89.1\%, 53.2\%, and 39.1\% for $n=4,6$, and $8$, respectively.
To the best of our knowledge, the 4-qubit result is by far the highest reported in any system and there have been no reports of generalized Toffoli gates with more than 4 qubits.
The relaxation-limited gate fidelities (total duration) for the 4-qubit, 6-qubit, and 8-qubit Toffoli gates are 92.5\% (0.4~$\mu$s), 66.7\% (1.3~$\mu$s, staggered pulses), and 62.3\% (1.1~$\mu$s), respectively, and are responsible for approximately 70\% of the total error; the remaining error is due in part to dephasing and in part to stray inter-qubit coupling \cite{chu2021coupler, cai2021impact, zajac2021spectator}.

\begin{figure*}[htbp]
	\centering
	\includegraphics[width=180mm]{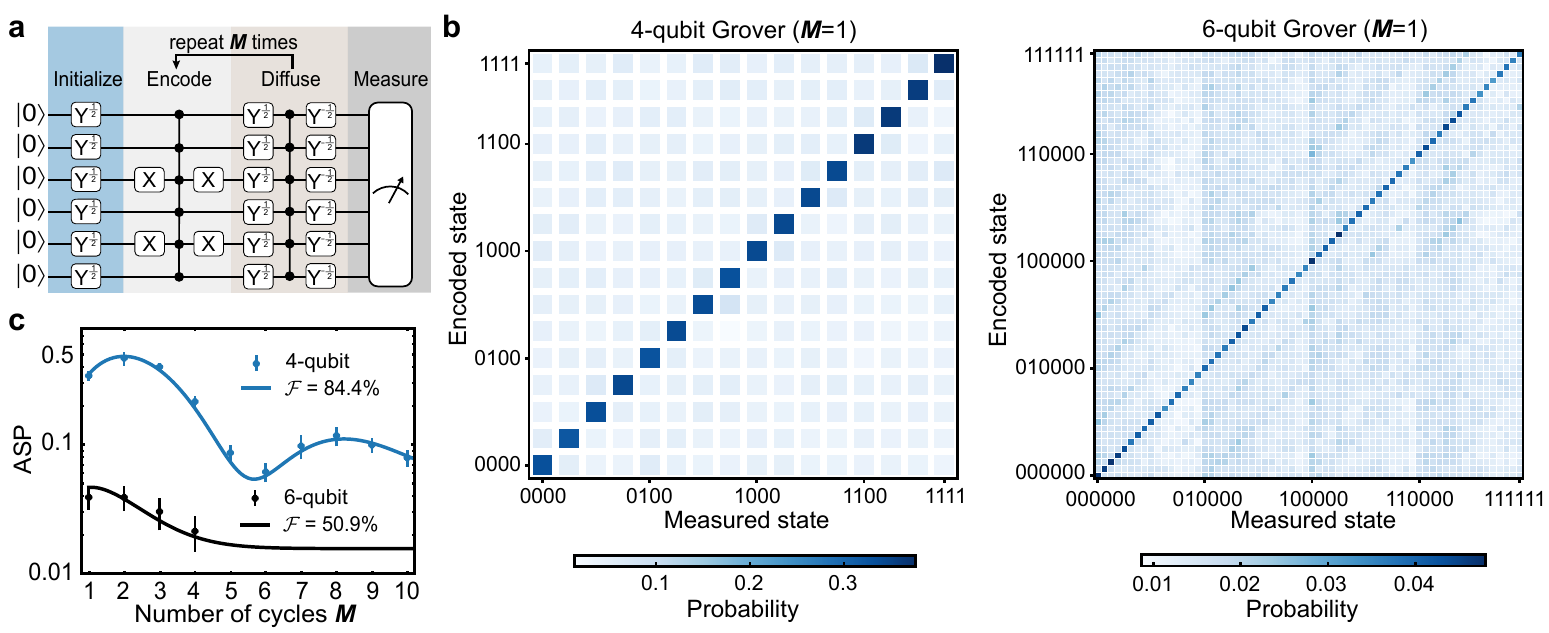}
	\caption{
		\textbf{Demonstration of Grover's search algorithm with multiple amplification cycles.}
		\textbf{a,} Circuit diagram implementing Grover's search algorithm. In this example, the encoded solution is 110101.
		$\rm{Y^{\pm1/2}}$ indicates a $\pm\pi/2$ rotation about the $Y$-axis.
		\textbf{b,} Measured output state probability distribution for each of the $2^n$ encoded states in the 4-qubit and 6-qubit Grover's search algorithms. 
		\textbf{c,} Average algorithm success probabilities (dots) and four times the standard deviations (error bars) of all $2^n$ cases versus the number of oracle-amplification cycles. 
		The solid lines are fit to Eq.~(\ref{eq:asp}) given finite gate fidelity $\mathcal{F}$.
	}
	\label{fig:grover}
\end{figure*}


Finally, we performed Grover's search algorithm as a complementary method to benchmark our multi-qubit gates. 
The core steps of this algorithm encode a solution bit-string $j$ with a phase oracle $O_{j} = \sum_{s \neq j} \ket{s}\bra{s} - \ket{j}\bra{j}$, a unitary that accesses the input function, and amplify the probability of finding \ket{j} via phase diffusion, with each step containing an $n$-qubit CZ gate (Fig.~\ref{fig:grover}a); these two steps may be repeated for further amplification.
Note that the phase oracle performs a conditional phase flip on \ket{j}; therefore, an arbitrary oracle can be constructed from an $n$-qubit CZ gate with additional pairs of $X$ gates applied to qubits being conditioned on \ket{0} instead of \ket{1}.

Figure \ref{fig:grover}b shows the results of the 4-qubit and 6-qubit single-solution Grover's search algorithms with one oracle-amplification cycle (see \cite{supplement} for extended data of the multi-solution Grover’s search).
The diagonal matrix elements correspond to the probabilities of finding the correct states, i.e., the algorithm success probability (ASP), and are substantially higher than the other elements, on average 34.2\% versus 4.4\% for the 4-qubit Grover’s search and 3.9\% versus 1.5\% for the 6-qubit Grover’s search, showing the effectiveness of the amplification. 
Because of its insufficient fidelity, the 8-qubit Grover result (not shown) does not display a significant ASP gain. 

To optimize ASP, we tested Grover's search algorithm with multiple rounds of amplification.
As shown in Fig.~\ref{fig:grover}c, the average ASP in the 4-qubit case shows a clear improvement to 46.8\% with one additional cycle ($M=2$), and a clear dependence is visible up to 10 cycles, that is, a total of 20 CCCZ gates, owing to the high gate fidelity.
Ignoring contributions from the single-qubit gate error, which is estimated to be 0.14\% from simultaneous randomized benchmarking, we developed a simplified model for estimating ASP \cite{supplement}:
\begin{equation}
{\rm ASP} = \mathcal{F}^{2M} \sin^2 \left( (2M+1) \arcsin( 2^{-\frac{n}{2}} ) \right) + \frac{1 - \mathcal{F}^{2M}}{2^n},
\label{eq:asp}
\end{equation}
where $\mathcal{F}$ is the $n$-qubit CZ gate fidelity.
Fitting the data to Eq.~(\ref{eq:asp}) gives $\mathcal{F}=$ 84.4\% and 50.9\% for the 4-qubit and 6-qubit cases, respectively, which are close to the above-measured truth-table fidelities.


The low-depth circuit synthesis using novel QuAND logic enabled our implementation of multiply controlled gates and Grover's search algorithm at record scale, confirming the feasibility of a scalable and resource-efficient approach to simplify algorithm compilation.
This study should not only stimulate interest in exploring alternative compilation schemes using QuAND logic but also help reduce hardware-related challenges, in particular, the connectivity problem for which solid-state devices has long been criticized.
Our work marks an essential step toward closing the gap between most anticipated near-term applications and available NISQ devices.

\textbf{Acknowledgement:} We thank Song Liu, Jingjing Niu, Youpeng Zhong, Orkesh Nurbolat and Sai Li for technical assistance and fruitful discussions.
This work was supported by 
the Key-Area Research and Development Program of Guang-Dong Province (Grant No. 2018B030326001), 
the National Natural Science Foundation of China (U1801661), 
the Guangdong Innovative and Entrepreneurial Research Team Program (2016ZT06D348), 
the Guangdong Provincial Key Laboratory (Grant No.2019B121203002), 
the Natural Science Foundation of Guangdong Province (2017B030308003), 
the Science, Technology and Innovation Commission of Shenzhen Municipality (KYTDPT20181011104202253), 
the Shenzhen-Hong Kong Cooperation Zone for Technology and Innovation (HZQB-KCZYB-2020050),
and the NSF of Beijing (Grant No. Z190012).
X.H. and X.S. acknowledge support from
the National Natural Science Foundation of China (Grant No. 61832003),
and the Strategic Priority Research Program of Chinese Academy of Sciences (Grant No. XDB28000000).

{\bf Author Contributions:}
J.C., X.H. and F.Y. conceived and designed the experiment. 
Y.Z. and F.Y. designed the device. 
J.C. conducted the measurements.
J.C., X.H., and J.Y. analyzed the data. 
Y.Z., H.J., and L.Z. performed sample fabrication. 
J.C., X.H., X.S. and F.Y. wrote the manuscript. 
X.S. and F.Y. supervised the project. 
All authors discussed the results and contributed to revising the manuscript and the supplementary information. 
All authors contributed to the experimental and theoretical infrastructure to enable the experiment.

\textbf{Data availability:}
The data that support the plots within this paper and other findings of this study are available from the corresponding authors upon reasonable request.

\bibliography{BibDoi}

\end{document}


\title{Supplementary Information for "Scalable algorithm simplification using quantum AND logic"}

\newcommand{\SIQSE}{\affiliation{1}{Shenzhen Institute for Quantum Science and Engineering (SIQSE), Southern University of Science and Technology, Shenzhen, Guangdong, China}}
\newcommand{\SIQA}{\affiliation{2}{International Quantum Academy, Shenzhen, Guangdong, China}}
\newcommand{\GDKL}{\affiliation{3}{Guangdong Provincial Key Laboratory of Quantum Science and Engineering, Southern University of Science and Technology, Shenzhen, Guangdong, China}}
\newcommand{\DPHY}{\affiliation{4}{Department of Physics, Southern University of Science and Technology, Shenzhen, Guangdong, China}}
\newcommand{\ICT}{\affiliation{5}{Institute of Computing Technology, Chinese Academy of Sciences, Beijing, China}}
\newcommand{\UCAS}{\affiliation{6}{University of Chinese Academy of Sciences, Beijing, China}}

\author{Ji Chu}
\thanks{These authors have contributed equally to this work.}
\affiliation{\SIQSE}\affiliation{\SIQA}\affiliation{\GDKL}
\author{Xiaoyu He}
\thanks{These authors have contributed equally to this work.}
\affiliation{\ICT}\affiliation{\UCAS}

\author{Yuxuan Zhou}
\affiliation{\SIQSE}\affiliation{\SIQA}\affiliation{\GDKL}\affiliation{\DPHY}
\author{Jiahao Yuan}
\affiliation{\SIQSE}\affiliation{\SIQA}\affiliation{\GDKL}\affiliation{\DPHY}
\author{Libo Zhang}
\affiliation{\SIQSE}\affiliation{\SIQA}\affiliation{\GDKL}

\author{Qihao Guo}
\affiliation{\SIQSE}\affiliation{\SIQA}\affiliation{\GDKL}\affiliation{\DPHY}
\author{Yong-ju Hai}
\affiliation{\SIQSE}\affiliation{\SIQA}\affiliation{\GDKL}\affiliation{\DPHY}
\author{Zhikun Han}
\affiliation{\SIQSE}\affiliation{\SIQA}\affiliation{\GDKL}
\author{Chang-Kang Hu}
\affiliation{\SIQSE}\affiliation{\SIQA}\affiliation{\GDKL}
\author{Wenhui Huang}
\affiliation{\SIQSE}\affiliation{\SIQA}\affiliation{\GDKL}\affiliation{\DPHY}
\author{Hao Jia}
\affiliation{\SIQSE}\affiliation{\SIQA}\affiliation{\GDKL}
\author{Dawei Jiao}
\affiliation{\SIQSE}\affiliation{\SIQA}\affiliation{\GDKL}
\author{Yang Liu}
\affiliation{\SIQSE}\affiliation{\SIQA}\affiliation{\GDKL}
\author{Zhongchu Ni}
\affiliation{\SIQSE}\affiliation{\SIQA}\affiliation{\GDKL}\affiliation{\DPHY}
\author{Xianchuang Pan}
\affiliation{\SIQSE}\affiliation{\SIQA}\affiliation{\GDKL}
\author{Jiawei Qiu}
\affiliation{\SIQSE}\affiliation{\SIQA}\affiliation{\GDKL}\affiliation{\DPHY}
\author{Weiwei Wei}
\affiliation{\SIQSE}\affiliation{\SIQA}\affiliation{\GDKL}
\author{Zusheng Yang}
\affiliation{\SIQSE}\affiliation{\SIQA}\affiliation{\GDKL}
\author{Jiajian Zhang}
\affiliation{\SIQSE}\affiliation{\SIQA}\affiliation{\GDKL}\affiliation{\DPHY}
\author{Zhida Zhang}
\affiliation{\SIQSE}\affiliation{\SIQA}\affiliation{\GDKL}\affiliation{\DPHY}
\author{Wanjing Zou}
\affiliation{\SIQSE}\affiliation{\SIQA}\affiliation{\GDKL}

\author{Yuanzhen Chen}
\affiliation{\SIQSE}\affiliation{\SIQA}\affiliation{\GDKL}\affiliation{\DPHY}
\author{Xiaowei Deng}
\affiliation{\SIQSE}\affiliation{\SIQA}\affiliation{\GDKL}
\author{Xiu-hao Deng}
\affiliation{\SIQSE}\affiliation{\SIQA}\affiliation{\GDKL}
\author{Ling Hu}
\affiliation{\SIQSE}\affiliation{\SIQA}\affiliation{\GDKL}
\author{Jian Li}
\affiliation{\SIQSE}\affiliation{\SIQA}\affiliation{\GDKL}
\author{Dian Tan}
\affiliation{\SIQSE}\affiliation{\SIQA}\affiliation{\GDKL}
\author{Yuan Xu}
\affiliation{\SIQSE}\affiliation{\SIQA}\affiliation{\GDKL}
\author{Tongxing Yan}
\affiliation{\SIQSE}\affiliation{\SIQA}\affiliation{\GDKL}

\author{Xiaoming Sun}
\email{sunxiaoming@ict.ac.cn}
\affiliation{\ICT}\affiliation{\UCAS}
\author{Fei Yan}
\email{yanf7@sustech.edu.cn}
\affiliation{\SIQSE}\affiliation{\SIQA}\affiliation{\GDKL}
\author{Dapeng Yu}
\affiliation{\SIQSE}\affiliation{\SIQA}\affiliation{\GDKL}\affiliation{\DPHY}

\date{\today}

\maketitle

\newpage

\tableofcontents

\section{Logic circuit construction using QuAND}

\begin{figure}[htbp] \subfloat[]{
	{
		\Qcircuit @C=1em @R=1em {
			\lstick{\ket{q_0}} &\ctrl{1}&\qw & &&& &\qw\qad&\qw&\qaw &\qw &\qw&\qw\\
			\lstick{\ket{q_1}} &\ctrl{1}&\qw & &&& &\gate{\&}&\qw\qad&\qw &\qw &\gate{\&}\qau&\qw\\
			\lstick{\ket{q_2}} &\ctrl{1}&\qw & & && &\qw&\gate{\&}&\ctrl{1} &\gate{\&}\qau &\qw&\qw\\
			\lstick{\ket{q_3}} &\multigate{2}{U}&\qw & &=&& &\qw&\qw&\multigate{2}{U} &\qw &\qw&\qw\\
			\lstick{\ket{q_4}} &\ghost{U}&\qw & & && &\qw&\qw&\ghost{U} &\qw &\qw &\qw\\
			\lstick{\ket{q_5}} &\ghost{U}&\qw & &&& &\qw&\qw &\ghost{U}&\qw &\qw &\qw \\\\
		}
	}
	}
	\vspace{.2in}
	
	\subfloat[]{
	{
		\Qcircuit @C=0.7em @R=1.2em {
			\lstick{\ket{q_0}} &\ctrl{1}&\qw & &&& &\qw &\ctrl{1} &\qw&\qw \\
			\lstick{\ket{q_1}} &\ctrl{1}&\qw & &&& &\qw &\ctrl{1} &\qw&\qw\\
			\lstick{\ket{q_2}} &\ctrl{1}&\qw & &&& &\qw &\ctrl{1} &\qw&\qw\\
			\lstick{\ket{q_3}} &\ctrl{1}&\qw & &=&&&\qw &\ctrl{1} &\qw&\qw&\\
			\lstick{\ket{q_4}} &\ctrl{1}&\qw & &&& &\qw &\ctrl{1} &\qw&\qw\\
			\lstick{\ket{q_5}} &\targ   &\qw & &&& &\gate{H} &\ctrl{-1} &\gate{H}&\qw\\\\
		}
	}
	}	
	\hspace{1in}
	\subfloat[]{
	{
		\Qcircuit @C=0.7em @R=1.2em {
			\lstick{\ket{q_0}} &\ctrl{1}&\qw & &&& &\qw &\ctrl{1} &\qw&\qw \\
			\lstick{\ket{q_1}} &\ctrl{1}&\qw & &&& &\qw &\ctrl{1} &\qw&\qw\\
			\lstick{\ket{q_2}} &\ctrl{1}&\qw & &&& &\qw &\ctrl{1} &\qw&\qw\\
			\lstick{\ket{q_3}} &\ctrl{1}&\qw & &=&& &\qw &\ctrl{1} &\qw&\qw&\\
			\lstick{\ket{q_4}} &\qswap  &\qw & &&& &\targ &\ctrl{1} &\targ&\qw\\
			\lstick{\ket{q_5}} &\qswap\qwx  &\qw & &&& &\ctrl{-1} &\targ &\ctrl{-1}&\qw\\\\
		}
	}
	}	
	\caption{\textbf{Circuit decomposition of generalized Fredkin gate.} \textbf{(a)} A circuit decomposition of multi-qubit controlled-$U$ gate (left) using using QuAND, reversal QuAND and controlled-$U$ gates (right).\textbf{(b)} A circuit decomposition of multi-qubit Toffoli gate (left) using $n$-qubit CZ and Hadamard gates (right).\textbf{(c)} A circuit decomposition of generalized Fredkin gate (left) using $n$-qubit Toffoli and CNOT gates.}
	\label{fig:fredkin}
\end{figure}
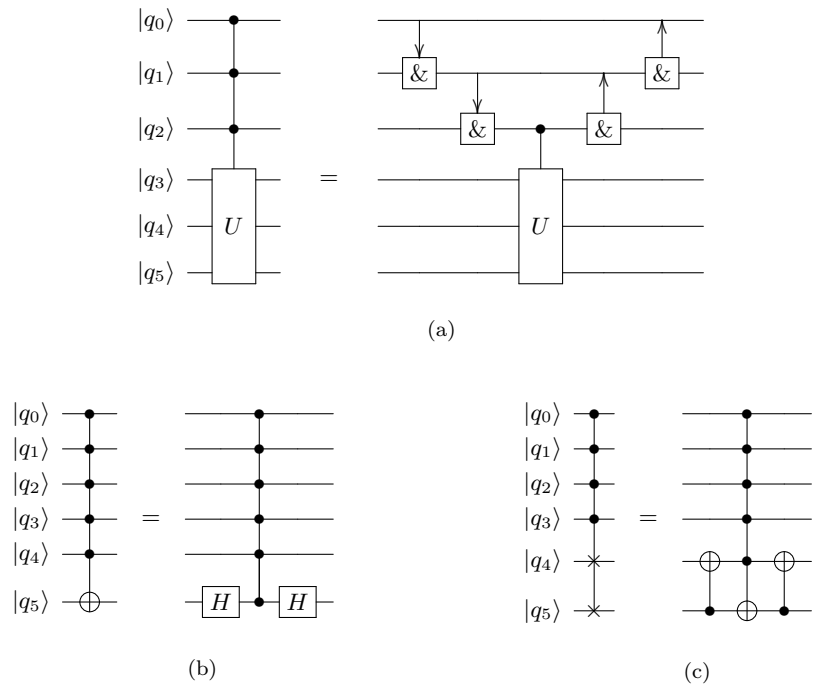

In the main text, we have shown the efficient decomposition for multi-qubit controlled-$Z$ ($n$-CZ) using the QuAND gate, which extends the classical AND logic to qubits. Basically, any multi-qubit controlled-unitary (CU) gates can be implemented efficiently in a same three-step procedure: embedding, controlled-unitary, and recovery (Fig.~\ref{fig:fredkin}a). 
Figure \ref{fig:fredkin}b and \ref{fig:fredkin}c also show alternative way to synthesize generalized Toffoli and Fredkin gate (controlled-SWAP), which are important quantum logics.

Since our QuAND gate is a quantum implementation of AND logic leveraging ancilla level and since NAND gate is universal in classical circuit, all classical logic circuits can be efficiently constructed by adapting classical circuit optimization techniques with single-qubit, two-qubit CZ and QuAND gates. In fact, compared to the traditional Toffoli decomposition scheme, our scheme requires fewer ancilla qubits and gate operations. 
The QuAND gate is readily applicable to a large category of circuits and useful in simplifying circuit synthesis.
Here we show three examples of leveraging QuAND gates for efficient synthesis of basic arithmetic circuits. 

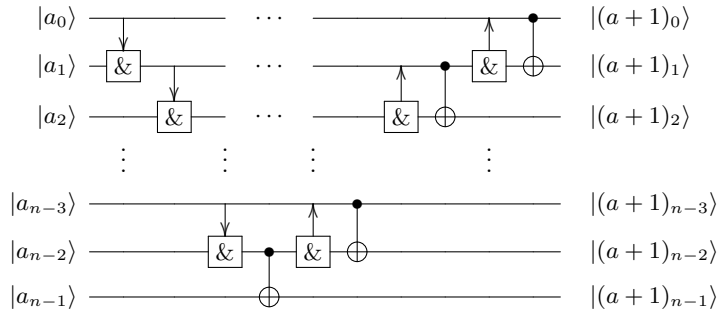
\begin{figure}[htbp] 
	\centerline{
		\Qcircuit @C=0.7em @R=0.8em {
			\lstick{\ket{a_0}} &\qw\qad &\qw      &\qaw &\cdots & &\qw&\qw&\qw&\qw &\ctrl{1} &\qw&\rstick{\ket{(a+1)_{0}}}\\
			\lstick{\ket{a_1}} &\gate{\&} &\qw\qad &\qw &\cdots & &\qw &\qw &\ctrl{1}&\gate{\&}\qau &\targ &\qw&\rstick{\ket{(a+1)_{1}}}\\
			\lstick{\ket{a_2}} &\qw    &\gate{\&} &\qw &\cdots & &\qw  &\gate{\&}\qau &\targ &\qw&\qw &\qw&\rstick{\ket{(a+1)_{2}}}\\
			& \vdots        &     &      \vdots  &       & \vdots   &&&&\vdots \\   \\
			\lstick{\ket{a_{n-3}}} &\qw &\qw &\qw\qad &\qaw      &\qw &\ctrl{1} &\qw&\qw&\qw &\qw&\qw&\rstick{\ket{(a+1)_{n-3}}}\\  
			\lstick{\ket{a_{n-2}}} &\qw &\qw &\gate{\&}     &\ctrl{1} &\gate{\&}\qau &\targ &\qw &\qw&\qw&\qw &\qw&\rstick{\ket{(a+1)_{n-2}}}\\  
			\lstick{\ket{a_{n-1}}} &\qw &\qw &\qw   &\targ &\qw &\qw  &\qw&\qw&\qw&\qw&\qw &\rstick{\ket{(a+1)_{n-1}}}\\
		}
	}
	\caption{\textbf{A circuit decomposition of incrementer using QuAND.} The $n$-qubit binary input $\ket{a}=\ket{a_{n-1}\dots a_{1}a_{0}}$ is incremented to $\ket{a+1}$ at the output. The subscript indicates the index of the binary digit.}
	\label{fig:inc}
\end{figure}

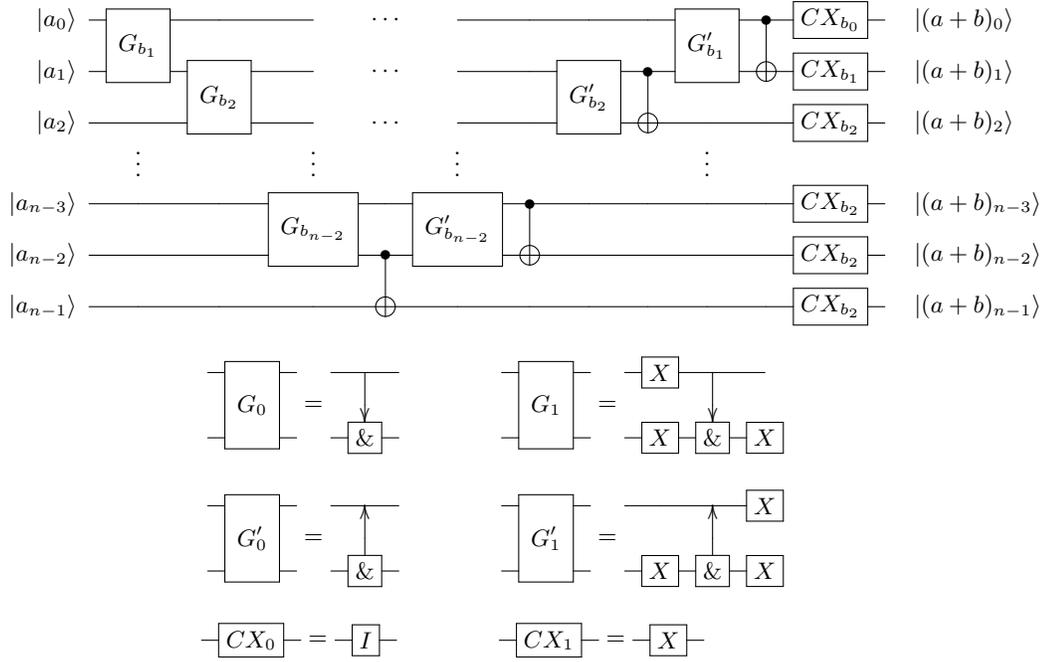
\begin{figure}[htbp] 
	\centerline{
		\Qcircuit @C=0.7em @R=.6em {
			\lstick{\ket{a_0}} &\multigate{1}{G_{b_1}} &\qw      &\qw &\cdots & &\qw&\qw&\qw&\multigate{1}{G_{b_1}'} &\ctrl{1} &\gate{CX_{b_0}}&\qw&\rstick{\ket{(a+b)_{0}}}\\
			\lstick{\ket{a_1}} &\ghost{G_{b_1}} &\multigate{1}{G_{b_2}} &\qw &\cdots & &\qw &\multigate{1}{G_{b_2}'} &\ctrl{1}&\ghost{G_{b_1}} &\targ &\gate{CX_{b_1}}&\qw&\rstick{\ket{(a+b)_{1}}}\\
			\lstick{\ket{a_2}} &\qw    &\ghost{G_{b_2}} &\qw &\cdots & &\qw  &\ghost{G_{b_2}} &\targ &\qw&\qw &\gate{CX_{b_2}}&\qw&\rstick{\ket{(a+b)_{2}}}\\
			& \vdots        &     &      \vdots  &       & \vdots   &&&&\vdots \\ \\  
			\lstick{\ket{a_{n-3}}} &\qw &\qw &\multigate{1}{G_{b_{n-2}}} &\qw      &\multigate{1}{G_{b_{n-2}}'} &\ctrl{1} &\qw&\qw&\qw &\qw&\gate{CX_{b_2}}&\qw&\rstick{\ket{(a+b)_{n-3}}}\\  
			\lstick{\ket{a_{n-2}}} &\qw &\qw &\ghost{G_{b_{n-2}}}     &\ctrl{1} &\ghost{G_{b_{n-2}}} &\targ &\qw &\qw&\qw&\qw &\gate{CX_{b_2}}&\qw&\rstick{\ket{(a+b)_{n-2}}}\\  
			\lstick{\ket{a_{n-1}}} &\qw &\qw &\qw   &\targ &\qw &\qw  &\qw&\qw&\qw&\qw&\gate{CX_{b_2}}&\qw &\rstick{\ket{(a+b)_{n-1}}}\\\\\\
		}
	} 
	
	\centerline{
		\Qcircuit @C=0.7em @R=0.7em {
			&\multigate{2}{G_0}&\qw    &  &&\qw &\qw & &&&&& &\multigate{2}{G_1}&\qw    &  &&\gate{X} &\qw &\qw\\
			&                  &  & =& &\qad\qwx         &  &&&&& &                  &  & &=& &         & \qad\qwx \\
			&\ghost{G_0}&\qw&       &&\gate{\&}      &\qw & &&&&&
			&\ghost{G_0}&\qw&       &&\gate{X}      &\gate{\&} &\gate{X}\\\\\\
		}
	}
	
	\centerline{
		\Qcircuit @C=0.7em @R=0.7em {
			&\multigate{2}{G_0'}&\qw    &  &&\qw &\qw & &&&&& &\multigate{2}{G_1'}&\qw    &  &&\qw &\qw &\gate{X}\\
			&                  &  & =& &\qau         &  &&&&& &                  &  & &=& &         & \qau \\
			&\ghost{G_0}&\qw&       &&\gate{\&}\qwx      &\qw & &&&&&
			&\ghost{G_0}&\qw&       &&\gate{X}      &\gate{\&}\qwx &\gate{X}\\\\\\
		}
	}

	\centerline{
		\Qcircuit @C=0.7em @R=0.7em {
			&\gate{CX_0}&\qw & = &&\gate{I}&\qw  & &&&&& &\gate{CX_1} &\qw & = &&\gate{X}&\qw &&&&&\\\\
		}
	}
	\caption{\textbf{A circuit decomposition of constant adder using QuAND.} The $n$-qubit binary input $\ket{a}=\ket{a_{n-1}\dots a_{1}a_{0}}$ is added by a known constant integer $b=b_{n-1}\dots b_{1}b_{0}$ ($b_0=1$) at the output.}
	\label{fig:constadd}
\end{figure}

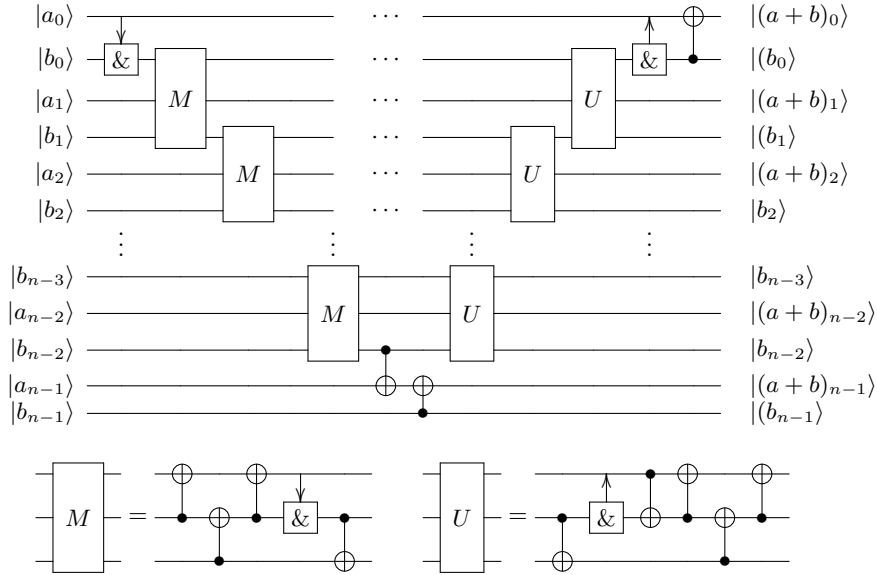
\begin{figure}[htbp] 
    \centerline{
		\Qcircuit @C=0.7em @R=0.6em {
		\lstick{\ket{a_0}} &\qw\qad &\qw &\qaw &\qw&\qw &\cdots  & &\qw &\qw  &\qw&\qw&\targ&\qw& \rstick{\ket{(a+b)_0}} \\
		\lstick{\ket{b_0}} &\gate{\&}&\multigate{2}{M} &\qw&\qw  &\qw &\cdots &&\qw &\qw&\multigate{2}{U}&\gate{\&}\qau&\ctrl{-1}&\qw&\rstick{\ket{(b_{0}}}\\
		\lstick{\ket{a_1}} &\qw         &\ghost{M} &\qw    &\qw &\qw&\cdots &&\qw &\qw&\ghost{U}  &\qw&\qw&\qw& \rstick{\ket{(a+b)_1}}\\
		\lstick{\ket{b_1}} &\qw&\ghost{M} &\multigate{2}{M}&\qw &\qw&\cdots &&\qw &\multigate{2}{U}&\ghost{U}   &\qw&\qw&\qw&\rstick{\ket{(b_{1}}}\\
		\lstick{\ket{a_2}} &\qw     &\qw         &\ghost{M}&\qw &\qw&\cdots &&\qw &\ghost{U} &\qw&\qw&\qw &\qw& \rstick{\ket{(a+b)_{2}}}\\
		\lstick{\ket{b_2}} &\qw     &\qw         &\ghost{M}&\qw &\qw&\cdots &&\qw &\ghost{U} &\qw&\qw&\qw&\qw&\rstick{\ket{b_2}}\\
		&\vdots&&&&\vdots &&&\vdots&&&\vdots \\
		\\
		\lstick{\ket{b_{n-3}}} &\qw &\qw &\qw &\qw &\multigate{2}{M} &\qw  &\qw   &\multigate{2}{U}&\qw&\qw&\qw&\qw&\qw&\rstick{\ket{b_{n-3}}}\\
		\lstick{\ket{a_{n-2}}} &\qw     &\qw  &\qw &\qw &\ghost{M} &\qw  &\qw  &\ghost{U}&\qw&\qw&\qw&\qw&\qw& \rstick{\ket{(a+b)_{n-2}}}\\
		\lstick{\ket{b_{n-2}}} &\qw     &\qw     &\qw &\qw &\ghost{M} &\ctrl{1}&\qw &\ghost{U}&\qw&\qw&\qw&\qw&\qw &\rstick{\ket{b_{n-2}}}\\
		\lstick{\ket{a_{n-1}}} &\qw   &\qw &\qw &\qw &\qw   &\targ  &\targ &\qw&\qw&\qw&\qw &\qw&\qw& \rstick{\ket{(a+b)_{n-1}}}\\
		\lstick{\ket{b_{n-1}}} &\qw   &\qw &\qw &\qw &\qw   &\qw    &\ctrl{-1}   &\qw &\qw&\qw&\qw&\qw&\qw   &\rstick{\ket{(b_{n-1}}}\\
		\\\\\\
		}
	}

	\centerline{
		\Qcircuit @C=0.7em @R=0.7em {
		&\multigate{2}{M}&\qw &  &&\targ  &\qw&\targ &\qw\qad &\qw&\qw &&&&\multigate{2}{U}&\qw &  &&\qw  &\qw &\ctrl{1} &\targ &\qw &\targ &\qw\\
		&\ghost{M}       &\qw &= &&\ctrl{-1}&\targ &\ctrl{-1}&\gate{\&}&\ctrl{1}&\qw&&& &\ghost{U}       &\qw &= &&\ctrl{1}&\gate{\&}\qau &\targ &\ctrl{-1}&\targ &\ctrl{-1}&\qw\\
		&\ghost{M}       &\qw &  &&\qw &\ctrl{-1}&\qw  &\qw  &\targ&\qw  &&&&\ghost{U}       &\qw &  &&\targ &\qw  &\qw  &\qw  &\ctrl{-1}&\qw&\qw
		}
	}

\caption{\textbf{A circuit decomposition of adder using QuAND.} Two $n$-qubit binary inputs, $\ket{a}=\ket{a_{n-1}\dots a_{1}a_{0}}$ and $\ket{b}=\ket{b_{n-1}\dots b_{1}b_{0}}$, are summed up at the output.
}
\label{fig:adder}
\end{figure}

Figure \ref{fig:inc} shows an efficient decomposition for incrementer using QuAND. The first half of the circuit -- a sequence of QuAND gates -- computes the carry information. The second half -- a sequence of CNOT and reverse QuAND gates -- recovers the original binary encoding and completes the incrementation. 
Figure \ref{fig:constadd} shows an efficient decomposition for constant adder using QuAND. The first half of the circuit -- a sequence of $G_0$ or $G_1$ (constructed by QuAND and single-qubit $X$ gates) gate depending on the corresponding bit value of $b$ -- computes the carry information. The second half of the circuit -- a sequence of reverse $G_0$ or reverse $G_1$, and CNOT gates -- recovers the original binary encoding and completes the addition operation. 
Figure \ref{fig:adder} shows an efficient decomposition for adder using QuAND, as inspired by \cite{cuccaro2004new}. The $M$ gate -- constructed by QuAND and CNOT gates -- computes the majority function and the carry information. The $U$ gates undo the $M$ gates and complete the addition of the two integers.

Depth of these circuits may be further reduced on topology with higher connectivity by the carry-lookahead technique.
Other arithmetic and boolean logic circuits can be constructed in a similar way that replaces AND gates in classical circuits with QuAND gates. 

\section{Multi-qubit Toffoli decomposition}
To compare different schemes for synthesizing $n$-qubit Toffoli gate, we list relevant references and their main properties in Table.\ref{table:alltoall} for all-to-all connection and in Table.~\ref{table:chain} for 1-D chain topology.

In prior works, multi-qubit Toffoli gate can be decomposed to qubit-only circuit with linear circuit depth and size by using ancilla qubits. 
The textbook approach \cite{nielsen2002quantum} reduces a big Toffoli to standard (3-qubit) Toffoli and costs $n-2$ ancilla qubits for concatenating the AND results. 
He et.al.\cite{he2017decompositions} provides a way to trade off between the number of ancilla qubits and circuit depth, but requires additional cost for feedback control or large constant factor. 
A similar approach from Barenco et.al.\cite{barenco1995elementary} uses the last control qubit as ancilla, saving ancilla qubit at the cost of circuit depth.
Other works have focused on simplifying $n$-qubit Toffoli by ancilla levels. 
Ralph et.al.\cite{ralph2007efficient} and Lanyon et.al.\cite{lanyon2009simplifying} utilizes $n$-level qudit system to achieve $2n$ circuit depth and size by swapping the target state out of qubit space when the control qubits are not $|1\rangle$.
Gokhale et.al.\cite{gokhale2019asymptotic} and Inada et.al.\cite{inada2021measurement} proposes leveraging qutrit control to achieve at most $2n$ circuit depth and size.
In particular, Gokhale et.al. \cite{gokhale2019asymptotic} proposes a novel approach which utilizes $|2\rangle$ state for storing the AND result of control qubits and propagate the results with Toffoli-like gate conditioned on $|2\rangle$ state, achieving logarithmic depth. 

In our scheme, the circuit depth and size are both in line with the best value from previous works. However, it requires only one additional operation with the ancilla level, i.e.\ the $\ket{11}\bra{20}+\ket{20}\bra{11}$ SWAP gate, which is naturally available in state-of-the-art hardware. The scheme features resource-efficient implementation in the sense that it is low-depth, free from ancilla qubits, and simple in control. These advantages plus compatibility with state-of-the-art hardware are the key to our successful realization of the large-scale multi-qubit Toffoli gate and Grover's search algorithm.

\begin{figure}[htbp]
	\centering
	\includegraphics[width=160mm]{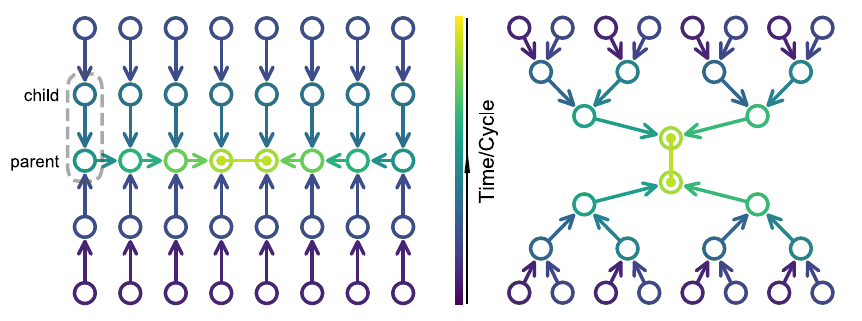}
	\caption{
		\textbf{Schematics of synthesizing multi-qubit CZ gates on 2-D square array (left) and a binary tree (right)}. Circles indicate qubits and arrows indicate QuAND gates, pointing from child to parent. Color gradient
		indicates the temporal order. The reverse QuAND sequence is omitted.
	}
	\label{fig:binaray_tree}
\end{figure}

Our scheme shows better scalability on qubit arrays with higher connectivity. The circuit depth can be reduced to 2$\sqrt{n}$ on a 2-D square array and to 2$\log_2 n$ on a binary tree, as shown in Fig.~\ref{fig:binaray_tree}.

\newpage
\begin{table*}[htbp]
	\begin{tabular}{c|c|c|c|c|c|c}
		&Depth         &Size     & Constant&Ancilla qubits &Control Requirement &Intuition
		\\\hline
		 Nielson and Chuang~\cite{nielsen2002quantum} & $\log_2 n$     & $n$     & $12$ & $n-2$  & Qubits &
		\begin{tabular}{c}
		$\quad\quad$
		\Qcircuit @C=0.7em @R=0.8em {
			\lstick{\ket{c_1}} &\ctrl{1} &\qw       &\ctrl{1} &\qw \\
			\lstick{\ket{c_2}} &\ctrl{1} &\qw       &\ctrl{1} &\qw \\
			\lstick{\ket{0}}   &\targ    &\ctrl{4}  &\targ &\qw \\
			\lstick{\ket{c_3}} &\ctrl{1} &\qw       &\ctrl{1} &\qw \\
			\lstick{\ket{c_4}} &\ctrl{1} &\qw       &\ctrl{1} &\qw \\
			\lstick{\ket{0}}   &\targ    &\ctrl{1}  &\targ &\qw \\
			\lstick{\ket{t}} &\qw      &\targ  &\qw &\qw \\
			\\
		}\end{tabular}
		\\\hline
		He et.al.~\cite{he2017decompositions} & $\log_2 n$     & $n$     & $4$ & $n-2$  & Qubits &
		\begin{tabular}{c}
		$\quad\quad$
		\Qcircuit @C=0.7em @R=0.8em {
			\lstick{\ket{c_1}} &\sctrl &\qw &\qw       &\ctrl{1} &\qw \\
			\lstick{\ket{c_2}} &\sctrl\qwx &\qw &\qw       &\ctrl{-1} &\qw \\
			\lstick{\ket{0}}   &\targ\qwx    &\ctrl{4}  &\gate{H}&\meter\cwx & \\
			\lstick{\ket{c_3}} &\sctrl &\qw&\qw       &\ctrl{1} &\qw \\
			\lstick{\ket{c_4}} &\sctrl\qwx &\qw&\qw       &\ctrl{-1} &\qw \\
			\lstick{\ket{0}}   &\targ\qwx    &\ctrl{1}  &\gate{H}&\meter\cwx & \\
			\lstick{\ket{t}}   &\qw      &\targ  &\qw &\qw &\qw \\
			\\
		}\end{tabular}
		
		\\\hline
		He et.al.~\cite{he2017decompositions} & $n$     & $n$     & $24$ & $1$  & Qubits &
		\begin{tabular}{c}
		$\quad\quad$
		\Qcircuit @C=0.4em @R=0.6em {
			\lstick{\ket{c_1}} &\ctrl{1} &\qw &\qw &\qw &\ctrl{1} &\qw &\qw &\qw\\
			\lstick{\ket{c_2}} &\ctrl{1} &\qw &\qw &\qw &\ctrl{1} &\qw &\qw &\qw\\
			\lstick{\ket{c_3}} &\ctrl{4} &\qw &\qw &\qw &\ctrl{4} &\qw &\qw &\qw\\
			\lstick{\ket{c_4}} &\qw      &\qw      &\ctrl{1} &\qw &\qw &\qw  &\ctrl{1} &\qw \\
			\lstick{\ket{c_5}} &\qw      &\qw      &\ctrl{1} &\qw &\qw &\qw  &\ctrl{1} &\qw \\
			\lstick{\ket{t}}   &\qw      &\gate{H} &\ctrl{1} &\qw &\qw &\qw  &\ctrl{1} &\gate{H} \\
			\lstick{\ket{a}}   &\targ    &\gate{S} &\targ &\gate{S^\dagger} &\targ &\gate{S}&\targ &\gate{S^\dagger} \\
			\\
		}\end{tabular}
		
		\\\hline
		 Barenco et.al.~\cite{barenco1995elementary} & $n^2$     & $n^2$     & $48$ & $0$  & Qubits &
		\begin{tabular}{c}
		$\quad\quad$
		\Qcircuit @C=0.0em @R=0.6em {
			\lstick{\ket{c_1}} &\qw &\ctrl{1} &\qw &\ctrl{1} &\ctrl{1} &\qw\\
			\lstick{\ket{c_2}} &\qw &\ctrl{1} &\qw &\ctrl{1} &\ctrl{1} &\qw\\
			\lstick{\ket{c_3}} &\qw &\ctrl{1} &\qw &\ctrl{1} &\ctrl{2} &\qw\\
			\lstick{\ket{c_4}} &\ctrl{1}         &\targ      &\ctrl{1}      &\targ &\qw  &\qw \\
			\lstick{\ket{t}}   &\gate{R_y(\pi/4)}&\qw &\gate{R_y(-\pi/4)}&\qw &\gate{R_y(\pi/4)}&\qw \\
			\\
		}\end{tabular}
		
		\\\hline
		Gokhale et.al.~\cite{gokhale2019asymptotic} & $\log_3 n$     & $n$     & $2$& $0$  & Three-Qutrit control &
		\begin{tabular}{c}
		$\quad\quad$
		\Qcircuit @C=0.7em @R=0.4em {
			\lstick{\ket{c_1}} &\onecontrol &\qw       &\onecontrol &\qw \\
			\lstick{\ket{c_2}} &\gate{X_{12}}\qwx &\twocontrol       &\gate{X_{12}}\qwx &\qw \\
			\lstick{\ket{c_3}} &\onecontrol\qwx &\qw\qwx &\onecontrol\qwx &\qw \\
			\lstick{\ket{c_4}} &\onecontrol &\qw\qwx       &\onecontrol &\qw \\
			\lstick{\ket{c_5}} &\gate{X_{12}}\qwx &\twocontrol\qwx       &\gate{X_{12}}\qwx &\qw \\
			\lstick{\ket{c_6}} &\onecontrol\qwx &\qw\qwx  &\onecontrol\qwx &\qw\\
			\lstick{\ket{t}}   &\qw             &\targ\qwx &\qw   &\qw\\
			\\
		}
		\end{tabular}
		\\\hline
		Ralph et.al.~\cite{lanyon2009simplifying,ralph2007efficient} & $n$     & $n$     & 2& 0  & n-level Qudit control &
		\begin{tabular}{c}
		$\quad\quad$
		\Qcircuit @C=0.4em @R=0.6em {
			\lstick{\ket{c_1}} &\qw     &\ctrlo{3}&\qw     &\qw     &\qw     &\ctrlo{3}&\qw    &\qw      \\
			\lstick{\ket{c_2}} &\qw     &\qw     &\ctrlo{2}&\qw     &\ctrlo{2}&\qw     &\qw    &\qw      \\
			\lstick{\ket{c_3}} &\qw     &\qw     &\qw    &\ctrl{1} &\qw     &\qw     &\qw     &\qw     \\
			\lstick{\ket{t}}   &\gate{H}&\gate{X_{12}}&\gate{X_{13}} &\ctrl{-1}   &\gate{X_{13}}&\gate{X_{12}}&\gate{H}&\qw     \\
			\\
		}\end{tabular}
		
		\\\hline
		This work & $\log_2 n$ & $n$ & 2& 0 & $\ket{11}\bra{20}+\ket{20}\bra{11}$ &
		\begin{tabular}{c}
		$\quad\quad$\Qcircuit @C=0.4em @R=0.4 em {
			\lstick{\ket{c_1}} &\qw\qad        &\qaw         &\qw &\qw &\qw &\qw \\
			\lstick{\ket{c_2}} &\gate{\&} &\qw         &\qw &\qw&\gate{\&}\qau &\qw \\
			\lstick{\ket{c_3}} &\qw \qad          &\qw\qwx\qad     &\qw &\qw\qau &\qw &\qaw\\
			\lstick{\ket{c_4}} &\gate{\&} &\gate{\&} &\ctrl{1} &\gate{\&}\qwx &\gate{\&}\qau &\qw\\
			\lstick{\ket{t}}   &\qw       &\qw       &\targ    &\qw  &\qw &\qw       \\
			\\
		}\end{tabular}
		
		\\\hline
	\end{tabular}
	\caption{\textbf{Comparison of multi-qubit Toffoli gate decomposition assuming all-to-all connectivity.}}
	\label{table:alltoall}
\end{table*}

\begin{table*}[htbp]
	\begin{tabular}{c|c|c|c|c|c|c}
		&Depth         &Size     & Constant&Ancilla qubits &Control Requirement &Intuition
		\\\hline
		Nielson and Chuang~\cite{nielsen2002quantum} & $n$     & $n$     & $>12$ & $n-2$  & Qubits &
		
		$\quad\quad$
		\Qcircuit @C=0.7em @R=0.8em {
			\lstick{\ket{c_1}} &\ctrl{1} &\qw     &\qw     &\qw    &\ctrl{1}\\
			\lstick{\ket{c_2}} &\ctrl{1} &\qw     &\qw     &\qw    &\ctrl{1} \\
			\lstick{\ket{0}}   &\targ    &\ctrl{1}&\qw     &\ctrl{1}&\targ \\
			\lstick{\ket{c_3}} &\qw      &\ctrl{1}&\qw     &\ctrl{1} &\qw \\
			\lstick{\ket{0}}   &\qw      &\targ   &\ctrl{1}&\targ &\qw\\
			\lstick{\ket{c_4}} &\qw      &\qw     &\ctrl{1}&\qw &\qw\\
			\lstick{\ket{t}}   &\qw      &\qw     &\targ   &\qw &\qw\\
			\\
		}
		\\\hline
		He et.al.~\cite{he2017decompositions} & $n$     & $n$     & $>4$ & $n-2$  & Qubits &
		$\quad\quad$
		\Qcircuit @C=0.7em @R=0.8em {
			\lstick{\ket{c_1}} &\sctrl    &\qw     &\qw       &\qw &\qw &\qw &\ctrl{1}&\qw\\
			\lstick{\ket{c_2}} &\sctrl\qwx &\qw    &\qw       &\qw &\qw &\qw &\ctrl{-1} &\qw\\
			\lstick{\ket{0}}   &\targ\qwx &\sctrl     &\qw     &\qw     &\ctrl{1}   &\gate{H}&\meter\cwx &\qw\\
			\lstick{\ket{c_3}} &\qw       &\sctrl\qwx &\qw     &\qw     &\ctrl{-1} &\qw &\qw&\qw\\
			\lstick{\ket{0}}   &\qw       &\targ\qwx  &\ctrl{1}&\gate{H}&\meter\cwx &\qw &\qw&\qw\\
			\lstick{\ket{c_4}} &\qw       &\qw        &\ctrl{1}&\qw&\qw&\qw&\qw&\qw\\
			\lstick{\ket{t}}   &\qw       &\qw        &\targ   &\qw &\qw &\qw &\qw&\qw\\
			\\
		}

		\\\hline
		Inada et.al.~\cite{inada2021measurement} & $n$     & $n$     & $2$& $0$  & Qutrit control &
		$\quad\quad$
		\Qcircuit @C=0.0em @R=0.4em {
			\lstick{\ket{c_1}} &\onecontrol       &\qw               &\qw&\qw&\qw&\qw&\onecontrol &\qw \\
			\lstick{\ket{c_2}} &\gate{X_{12}}\qwx &\twocontrol       &\qw&\qw&\qw&\twocontrol&\gate{X_{12}}\qwx &\qw \\
			\lstick{\ket{c_3}} &\qw               &\gate{X_{12}}\qwx &\twocontrol &\qw&\twocontrol&\gate{X_12}\qwx &\qw &\qw\\
			\lstick{\ket{c_4}}   &\qw               &\qw &\gate{X_{12}}\qwx   &\twocontrol&\gate{X_{12}}\qwx&\qw&\qw&\qw\\
			\lstick{\ket{t}}   &\qw               &\qw &\qw &\gate{X_{ij}}\qwx   &\qw&\qw&\qw&\qw\\
			\\
		}

		\\\hline
		This work & $n$ & $n$ & $2$& $0$ & $\ket{11}\bra{20}+\ket{20}\bra{11}$ &
		$\quad\quad$\Qcircuit @C=0.4em @R=0.6em {
			\lstick{\ket{c_1}} &\qw\qad   &\qw       &\qaw      &\qw           &\qw &\qw\\
			\lstick{\ket{c_2}} &\gate{\&} &\qw\qad   &\qw      &\qw           &\gate{\&}\qau&\qw\\
			\lstick{\ket{t}}   &\gate{H}  &\gate{\&} &\ctrl{1} &\gate{\&}\qau &\gate{H}     &\qw\\
			\lstick{\ket{c_3}} &\qw       &\gate{\&} &\ctrl{-1}&\gate{\&}\qad &\qw &\qw\\
			\lstick{\ket{c_4}} &\gate{\&} &\qw\qau   &\qw      &\qw           &\gate{\&}\qad &\qw\\
			\lstick{\ket{c_5}} &\qw\qau   &\qw       &\qaw      &\qw           &\qw &\qw\\
			\\
		}
		
		\\\hline
	\end{tabular}
	\caption{\textbf{Comparison of multi-qubit Toffoli gate decomposition assuming a 1-D chain.}}
	\label{table:chain}
\end{table*}

\section{Device and experimental setup}

\subsection{Wiring}

The processor is made of aluminum on sapphire following a similar recipe as described in Ref.~\cite{qiu2021suppressing}.
It is mounted inside a dilution refrigerator at a base temperature of 10~mK. We magnetically shield the processor with two Cryoperm cylinders. 
Inside the refrigerator, we use a total of 10 coaxial lines for the qubit/coupler control, 1 input and 1 output line for qubit readout. 
Attenuators and filters are installed at different temperature stages for thermalization and noise attenuation. 
At the lowest-temperature stage, we use customized low-pass filters for attenuating noise on all the control lines.  
The output signals are amplified by a high electron mobility transistor (HEMT) amplifier (40~dB gain) at the 4K stage and another low-noise amplifier (50~dB gain) at room temperature. Circulators and filters are placed on the output line to block noises from higher temperature stages. The output signals are finally down-converted to intermediate frequency and demodulated by two analog-to-digital converters (ADCs)

Microwave signals for single-qubit XY control and dispersive readout are up-converted from carriers generated by a microwave source using IQ mixing. We use diplexers to combine the XY and the Z signals at room temperature. For better impedance matching, an isolator is added to the XY port.

\subsection{Device parameters}

We summarize the measured device parameters in Table.\Rmnum{3}. Over time, we observe coherence fluctuations for som qubits, likely due to coupling to spurious two-level systems (TLSs).

We observe significant dephasing (Ramsey decay time <200~ns) from flux noise when performing two-qubit operations by adjusting the coupler frequency close to the qubit frequency. We fit the dephasing time to the slope of corresponding energy spectrum in order to extract the flux noise amplitude $\sigma_{\Phi_{\rm e}} = 116 \mu \Phi_0$ according to the relation $\Gamma_{\phi}^{\rm s}(\Phi_{\rm e}) = \frac{1}{\sqrt{2}} \frac{\partial \widetilde{\omega}_{\rm s}}{\partial \Phi_{\rm e}} \sigma_{\Phi_{\rm e}}$.
The relaxation time of the couplers are measured by swapping excitation back and forth between the coupler and its neighboring qubit.
The strong flux noise and shorter relaxation time of the couplers explain majority of the two-qubit gate error. 

We use randomized benchmarking (RB) to characterize gate errors. 
For single qubit gates, the gate fidelity in both isolated and simultaneous benchmarking is near the coherence limit. For two-qubit gates, we observe worse error rate -- typically 2-3 times -- when performing simultaneous two-qubit RB experiments on all eight qubits. We suppose the cause is the spectator effect in the presence of unwanted stray coupling as discussed in \cite{chu2021coupler}.

\begin{table}[h!]
	\begin{center}
		\begin{threeparttable}
			\begin{tabular}{|p{5.6cm}<{\centering}|p{1.3cm}<{\centering}|p{1.2cm}<{\centering}|p{1.2cm}<{\centering}|p{1.2cm}<{\centering}|p{1.2cm}<{\centering}|p{1.2cm}<{\centering}|p{1.2cm}<{\centering}|p{1.2cm}<{\centering}|}
				\hline
				\hline
				Qubit$^{\rm a}$ & $Q_0$  & $Q_1$ & $Q_2$ &  $Q_3$ & $Q_4$ &  $Q_5$ &  $Q_6$ &  $Q_7$\\
				\hline
				Frequency (GHz) & 6.257 & 7.015 & 6.359 & 7.152 & 6.424 & 7.142 & 6.504 & 7.184 \\
				Anharmonicity (GHz) & -0.255 & -0.236  & -0.252 & -0.239 & -0.252 & -0.238  & -0.250  & -0.239  \\
				Readout resonator frequency (GHz) & 4.973 & 4.979 & 5.018 & 5.023  & 5.064 & 5.071 & 5.112 & 5.120  \\
				Readout resonator linewidth (MHz) & 0.7 & 0.8 & 0.7 & 0.8 & 0.8 & 0.8 & 0.8 & 0.8 \\
				Dispersive shift of $|1\rangle$ (MHz) &  -0.55 & -0.22  & -0.50  & -0.22 & -0.47   & -1.20  & -0.50  & -0.28  \\
				Dispersive shift of $|2\rangle$ (MHz) &   -1.65 &  -0.75 &  -1.85 &  -0.67 &  -1.75  &  -0.60 &  -1.40 & -0.80 \\
				Readout error of $|0\rangle$ (\%) & 3.3 & 1.4  & 1.3 & 1.1& 1.1 & 3.9  & 4.0 & 1.2 \\
				Readout error of $|1\rangle$ (\%)  &  10.1 &  14.4  &  8.9 &  10.3 & 7.3 &  22.3  &  14.2  &  8.8  \\
				Relaxation time of $|1\rangle$  ($\mu s$) & 21 & 10 & 15  & 19 & 22 & 7.3 & 12 & 18  \\
				Relaxation time of  $|2\rangle$ ($\mu s$) & 14 &  6.1 &  8.0 &  10 &  12 & 4.6 & 7.1 & 8.9  \\			
				Ramsey decay time ($\mu s$) & 25 & 11 & 19 & 13 & 26 & 12 & 18 & 9.0  \\
				Spin echo decay time ($\mu s$) & 32 & 15 & 23 & 32 & 36 & 11 & 21 & 30 \\
				1-Q gate error$^{\rm c}$ (isol.)(\%) & 0.07 &  0.16 &  0.11 & 0.11 &  0.06 &  0.19  &  0.11 &  0.14  \\
				1-Q gate error$^{\rm c}$ (simul.)(\%) & 0.10 & 0.17 & 0.12 & 0.13 & 0.07 & 0.23 & 0.14 & 0.13 \\
				\hline
				\hline
				Coupler & $C_{01}$  &  $C_{12}$& $C_{23}$ &  $C_{34}$ &  $C_{45}$ &   $C_{56}$ &  $C_{67}$ &  $C_{70}$ \\
				\hline 
				Frequency (sweet spot) (GHz) & 8.08 & 8.13 & 8.04 & 8.28 & 8.38 & 8.27 & 8.40 & 8.32  \\
				Relaxation time ($\mu s$) & 4.6 & 4.9 & 7.5 & 6.7  &  4.5 & 7.1 & 3.1 & 3.3  \\
				Flux noise strength ($\mu\Phi_0$) & 118 & 120 & 116 & 117 &  97 & 108 & 144 & 110  \\
				Coupling coefficient$^{\rm b}$$r_{\rm qc}$  (left qubit)  & 0.0186  & 0.0189 & 0.0185 & 0.0070  & 0.0184  & 0.0185  & 0.0186 & 0.0061   \\
				Coupling coefficient$^{\rm b}$ $r_{\rm qc}$ (right qubit) &  0.0155 & 0.0175 & 0.0177 &   0.0098 &  0.0175 &  0.0176 & 0.0178 & 0.0076  \\
				CZ gate error$^{\rm c}$ (isol.)(\%) & 0.92 & 0.97 & 1.20 & 3.50 & 1.40 & 1.50 & 1.2 & 3.0  \\
				CZ gate error$^{\rm c}$ (simul.)(\%) & 2.5 & 5.0 & 3.4 & 7.5 & 3.0 & 5.1 & 2.4 & 6.0  \\
				\hline
				\hline		
			\end{tabular}
			\begin{tablenotes}
				\footnotesize
				\item[a] The qubit parameters are measured with the coupler idly biased at the sweet spot. 
				\item[b] The coupling strength between the qubits and the coupler is frequency dependent, as $g_{\rm qc} = r_{\rm qc}\sqrt{\omega_{\rm q} \omega_{\rm c}}$. We extract the coupling coefficient by fitting the measured level spectra versus the flux bias on the coupler.
				\item[c] The gate errors are measured using single(two)-qubit randomized benchmarking (RB). For two-qubit RB, we extract CZ gate errors by subtracting single qubit errors from the average errors of two-qubit Cliffords. The simultaneous two-qubit RB is performed in group of ($Q_0$-$Q_1$,$Q_2$-$Q_3$,$Q_4$-$Q_5$,$Q_6$-$Q_7$) and ($Q_1$-$Q_2$,$Q_3$-$Q_4$,$Q_5$-$Q_6$,$Q_7$-$Q_0$).
			\end{tablenotes}
		\end{threeparttable}
	\end{center}
	\caption{\textbf{Device parameters}.}
	\label{table:Device}
\end{table}	
	
\begin{figure}[htbp]
	\centering
	\includegraphics[width=160mm]{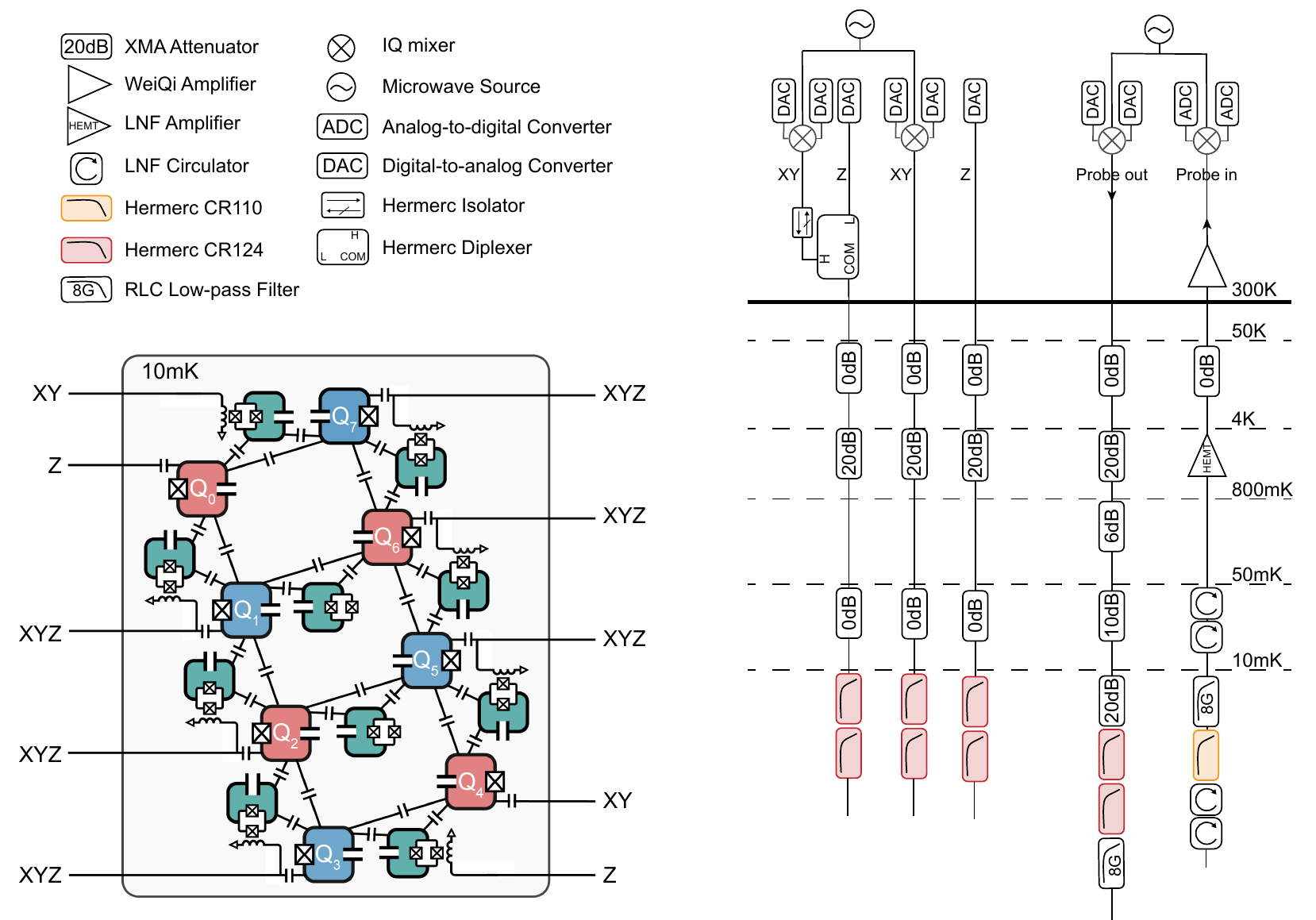}
	\caption{
		\textbf{Schematic diagram of the experimental setup}. (Left) Schematic diagram of the 8-qubit quantum processor and 10 control lines (the readout line and the resonators are not shown). (Right) The schematic diagram of control electronics and wiring.}
	\label{fig:wiring}
\end{figure}

\begin{figure}[htbp]
	\centering
	\includegraphics[width=140mm]{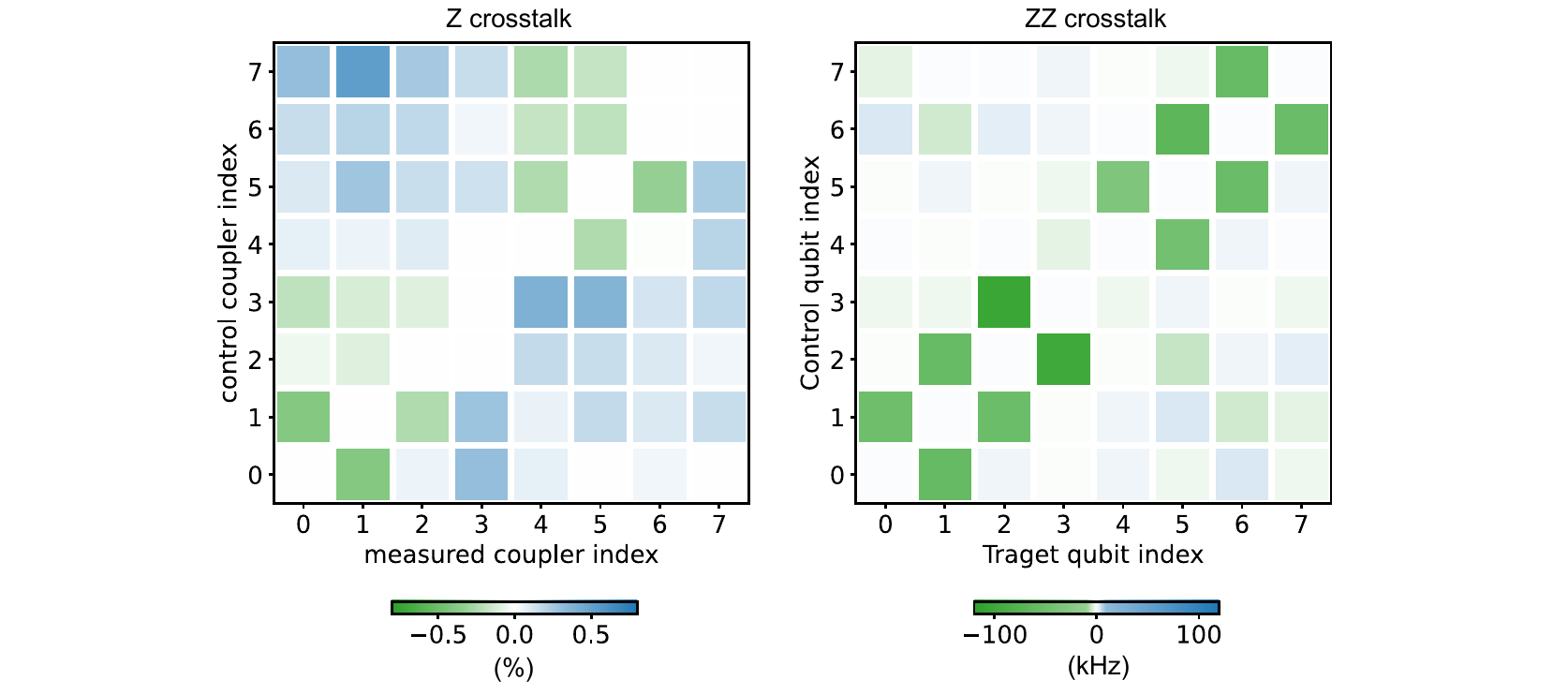}
	\caption{
		\textbf{Z crosstalk coefficient (left) between couplers and ZZ interaction strength (right) between qubits}. The ZZ interaction strength is measured with all the couplers biased at the sweet spot.}
	\label{fig:crosstalk}
\end{figure}

\subsection{Crosstalk}
Crosstalk is a major technical challenge for large-scale processors.
In Fig.~\ref{fig:crosstalk}, we show the measured flux (Z) crosstalk and residual $ZZ$ interaction.
The Z crosstalk is strongest between neighboring qubits (average: 0.23\%, standard deviation: 0.16\%).
The ZZ crosstalk is also strongest between neighboring qubits (average: 52~kHz, standard deviation: 26~kHz).
Both show very low level crosstalk.
In addition, the effect from microwave (XY) crosstalk between neighboring qubits (not shown) are negligible in this device due to large detuning between red-band and blue-band qubits. There is negligible effect from XY crosstalk as evident by isolated and simultaneous single-qubit gate RB results (Table \ref{table:Device}).

\begin{figure}[htbp]
	\centering
	\includegraphics[width=100mm]{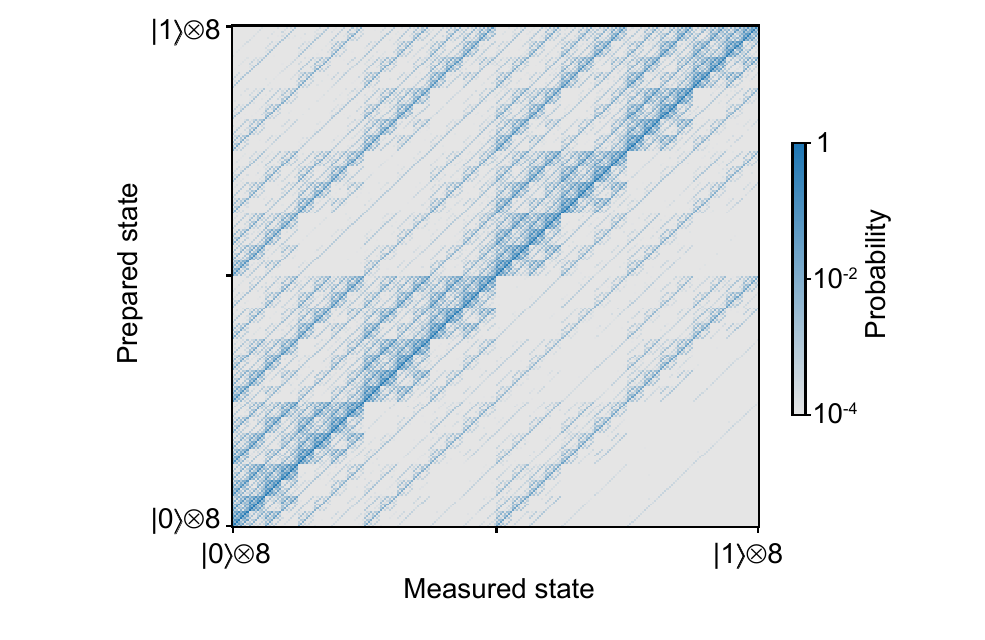}
	\caption{
		\textbf{Multi-qubit readout correction matrix}. Readout matrix for 8-qubit system $|Q_0Q_1Q_2Q_3Q_4Q_5Q_6Q_7\rangle$.The readout matrix is measured by traversing $2^8 = 256$ eigenstates of the system (lowest two states for each qubit).  
		For each prepared state, we repeat multiplexed state measurement for 50000 times.}
	\label{fig:readmat}
\end{figure}

\subsection{Readout correction}
To correct state preparation and measurement (SPAM) error, we first find the transfer matrix $\mathcal{R}$ by preparing all computational states of the joint system and measured the final probability distribution, as shown in Fig.~\ref{fig:readmat}. Given the relatively small single-qubit gate error (0.14\%), we have ignored the error of $\mathcal{R}$ itself. 
We then use $\mathcal{R}$ to correct readout results in subsequent experiments according to
\begin{equation}
	|\varphi_{\rm corrected}\rangle =  \mathcal{R}^{-1} |\varphi_{\rm raw}\rangle.
\end{equation}

\section{Coupler-assisted SWAP gate}

\begin{figure}[htbp]
	\centering
	\includegraphics[width=170mm]{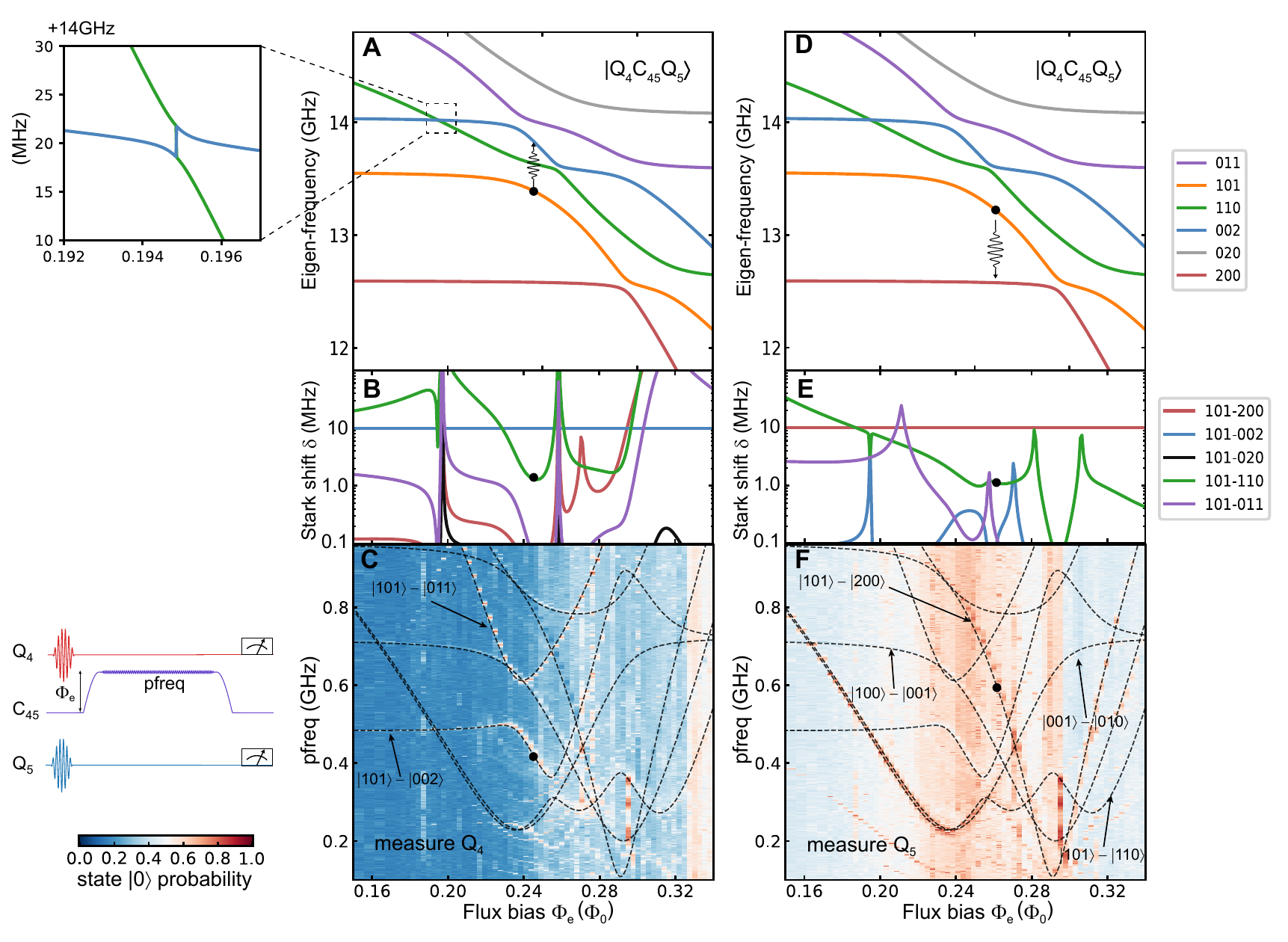}
	\caption{
		\textbf{Optimizing the operation point for the SWAP gate}.
		\textbf{(A)} Energy level spectra of the qubit(${\rm Q}_4$)-coupler-qubit(${\rm Q}_5$) system versus the coupler flux bias $\Phi_{\rm e}$ (solid lines). The inset shows the small avoided level crossing between \ket{002} and \ket{110}.
		\textbf{(B)} AC Stark shift induced by spurious transitions from the parametric drive versus flux bias. At each bias, we assume a drive amplitude that corresponds to 10~MHz \ket{101}-\ket{002} swapping rate, i.e.\ $\Omega=$10~MHz.
		Several prominent transitions are identified.
		\textbf{(C)} Transition spectroscopy (initial state: \ket{101}, measured qubit: $Q_4$) by sweeping the frequency of the parametric pulse and the the flux bias amplitude. The experiment sequence is shown in the left.
		Here we identify transitions by measuring the probability of \ket{0}, since the chosen readout frequency doesn't discriminate the first and second excited state.
		The black dots sharing the same horizontal axis indicate the optimal operation point for the parametric swap between \ket{101} and \ket{002}.
		\textbf{(D)} Same as (A) but the transition is to \ket{200}.
		\textbf{(E)} Same as (B) but the resonant transition is between \ket{101}-\ket{200}.
		\textbf{(F)} Same as (C) but the measured qubit is $Q_5$.
	}
	\label{fig:spectro}
\end{figure}

	\subsection{Theory}

	The tri-mode (Q-C-Q) system Hamiltonian in the laboratory frame under a flux drive $H_{\rm drive}(t)$ is ($\hbar=1$)
	\begin{equation}
	\begin{aligned}
	H & = H_{\rm static} + H_{\rm drive}(t), \quad \rm where\\
	H_{\rm static} &= \sum_{i=1,2,\rm c}(\omega_{i}a_i^{\dagger}a_i+\frac{\alpha_i}{2} a_i^{\dagger}a_i^{\dagger}a_ia_i) + \sum_{i=1,2} g_{i \rm c}( a_i^{\dagger}+a_i )( a_{\rm c}^{\dagger}+a_{\rm c} ) 
	+g_{12}( a_1^{\dagger}+a_1 )( a_2^{\dagger}+a_2 ) \quad \rm and \\
	H_{\rm drive} &= H_{\rm adia}(t) + H_{\rm dia}(t) \,. \label{eq:hamiltonian_full}
	\end{aligned} 
	\end{equation}
	The drive Hamiltonian can be divided into the adiabatic part $H_{\rm adia}(t)$ and the parametric drive $H_{\rm dia}(t) = \xi(t) a_{\rm c}^{\dagger}a_{\rm c} = A(t)\cos(\wdr t+\thdr) a_{\rm c}^{\dagger}a_{\rm c} $. Note that we have absorbed a drive-amplitude-dependent frequency shift from the nonlinear relation between coupler frequency and applied flux into the adiabatic part. 
	Following the instantaneous eigenbasis defined by $H_{\rm static}+H_{\rm adia}(t)$, we may rewrite an approximate Hamiltonian concerning only two levels of interest ($\ket{i}$ and $\ket{j}$),
	\begin{equation}
	\begin{aligned}
	H_{\rm TLS}(t) & = \frac{1}{2} \left[ \Delta_{ij} + \xi(t)\delta_{ij} \right] \hat{\sigma}_{\rm z} + \xi(t) (n_{ij} \hat{\sigma}_{-}  + \rm{h.c.} ) \\
	\end{aligned}
	\end{equation}
	where $\hat{\sigma}_{\rm z} =  |i\rangle \langle i | - |j\rangle \langle j|$, $\hat{\sigma}_{-} =  |i\rangle \langle j |$, $\Delta_{ij}$ is the instantaneous level spacing, $n_{ij} = \langle i |a_{\rm c}^{\dagger}a_{\rm c}| j \rangle $, and $\delta_{ij} = n_{ii} - n_{jj}$. 
	
	Defining the following unitary operator,
	\begin{equation}
	\Lambda(t) = e^{{\rm i} \hat{\sigma}_z [\Delta_{ij} t + \zeta(t) \delta_{ij} ] / 2},
	\end{equation}
	where $\zeta(t) = \int_{0}^{t} \xi(t')\rm{d}t'$ and assuming a constant drive amplitude $A(t)=b$, we can express the effective Hamiltonian in the rotating frame as
	\begin{equation}
	\begin{aligned}
	H'_{\rm TLS}(t) &= \Lambda H_{\rm TLS} \Lambda^{\dagger} + i \partial_{t} \Lambda \Lambda^{\dagger} \\
	& = \frac{1}{2} (\tilde{\Omega} \, \hat{\sigma}_{-} + \rm{h.c.} ),
	\end{aligned}
	\label{eq:hamiltonian}
	\end{equation}
	where $\tilde{\Omega} = b \left[ \J_0(\frac{b\delta_{ij}}{\wdr}) + \J_2(\frac{b\delta_{ij}}{\wdr}) \right] n_{ij} e^{\i [(\Delta_{ij} - \wdr)t - \thdr]} $.
	In the above equation, we have omitted fast oscillating terms and high order Bessel terms in the Jacobi-Anger expansion.
	
	Under the resonant condition $\Delta_{ij} - \omega_{\rm d} = 0$, the corresponding unitary operator in the subspace is 
	\begin{equation}
	U'_{\rm TLS} = \begin{pmatrix}
	& \cos(\Omega  t/2) & -\i \e^{-\i \thdr} \sin(\Omega t/2)\\
	& -\i\e^{\i \thdr} \sin(\Omega  t/2) & \cos( \Omega t/2) \,,\\
	\end{pmatrix}
	\label{eq:unitary}
	\end{equation}
	where $\Omega=|\tilde{\Omega}|$ and an irrelevant initial phase from $n_{ij}$ is ignored. The SWAP operation is realized by setting $ \Omega t  = \pi$ (or $\int \Omega(t) {\rm d}t = \pi$ for time-dependent envelope).
	
\begin{figure}[htbp]
	\centering
	\includegraphics[width=160mm]{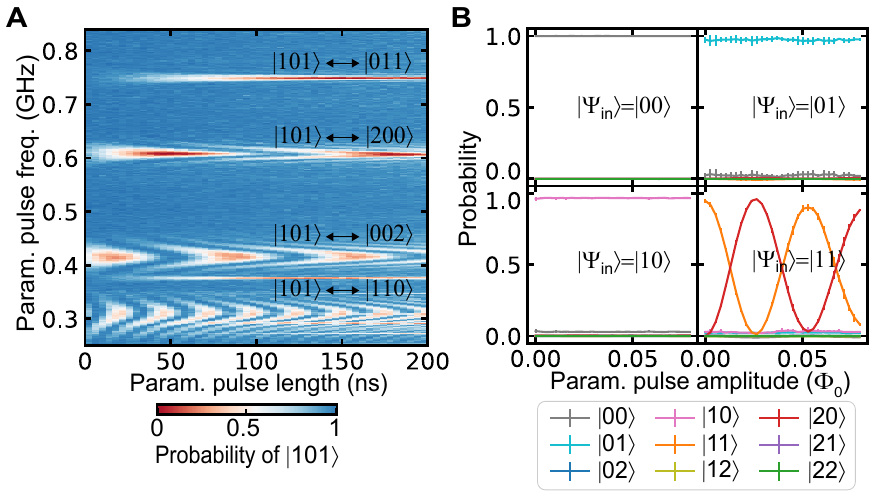}
	\caption{
		\textbf{Implementation of the coupler-assisted SWAP gate}.
		\textbf{(A)} Swap spectroscopy (starting from \ket{101})by sweeping the length and frequency of the parametric pulse at $\Phi_{\rm e} = 0.26~\Phi_0$, using a weak parametric pulse amplitude ($A_{\rm p} \approx 0.01$).
		Several prominent transitions are identified. 
		\textbf{(B)} Measured probability distribution of the final state after the SWAP pulse (coupler traced out) versus the parametric pulse amplitude, repeated for four different input states. The parametric pulse length is 30 ns.
	}
	\label{fig:swap}
\end{figure}	

	\subsection{Gate optimization}
	
	For optimal state transfer during the SWAP gate, the following four factors need to be considered when choosing the operation point: (1) short gate time considering the limited coherence time; (2) avoiding unwanted transitions during rise and fall; (3) smaller parametric modulation amplitude to avoid high-order interaction; (4) avoiding spurious transitions during the parametric modulation.
	
	Short gate time can be realized by shortening either the adiabatic rising/falling edges or the parametric pulse length. 
	To loosen the adiabatic condition and ease the pulse shaping efforts, operation points closer to the idling bias (towards the left in the Fig.~\ref{fig:spectro}A and D) are preferred.
	It's worth noting that there is generally a small avoided level crossing ($\Delta_{\rm gap} \approx 2$ MHz) between \ket{110} and \ket{002}, which we want to pass as fast as possible to avoid transition to \ket{002}. 
	Shortening the parametric pulse requires stronger parametric modulation amplitude and/or a larger transition matrix element $|\langle s |a_{\rm c}^{\dagger}a_{\rm c}| 101 \rangle|$ ($s$=200 or 002) according to Eq.~\ref{eq:hamiltonian}. 
	Too strong a parametric modulation results in high-order interaction term that cannot be ignored in the Jacobi-Anger expansion. For small modulation amplitude and short length, the working point towards lower coupler frequency is preferred as the transition matrix element becomes significantly stronger due to wavefunction hybridization.
	
	To avoid spurious transitions to other states, we compare their impact on the target transition across the bias range.
	Figure \ref{fig:spectro}B and E show the calculated AC Stark shift $\delta = (\sqrt{\Delta^2 + (r~\Omega_0)^2} - |\Delta|)$, a metric to quantify the spurious effect by taking into consideration both detuning from targeted to unwanted transition $\Delta$ and drive strength $r*\Omega_0$, where $\Omega_0$ is the Rabi frequency of the targeted transition given a certain drive amplitude, and $r$ is the relative strength of the unwanted transition.
 	In the experiments, we choose the extremum (black dot) as our operation point, balancing both the spurious effect and adiabatic constraint.
 	The working points can be roughly identified from the transition spectroscopy experiment by comparing theory and experimental result, as shown in Fig.~\ref{fig:spectro}B and E.

	At the selected operation points, we identify \ket{101}$\leftrightarrow$\ket{200} transition through a swap-spectroscopy experiment (Fig.~\ref{fig:swap}A). The targeted transition is well-separated from other spurious transitions.
	Finding the parametric frequency, we calibrate the SWAP gate (pulse width: 30~ns) by sweeping the pulse amplitude, as shown in Fig.~\ref{fig:swap}B.
	We check the integrity of the selected transition by counting the probability at each computational state, and repeat it for four different input states.
	Residual transition errors (2.7\% on average) are mainly caused by energy relaxation during the pulse.

\section{Implementation of $n$-qubit CZ gates}

\subsection{Gate decomposition}

\begin{figure}[htbp]
	\centering
	\includegraphics[width=180mm]{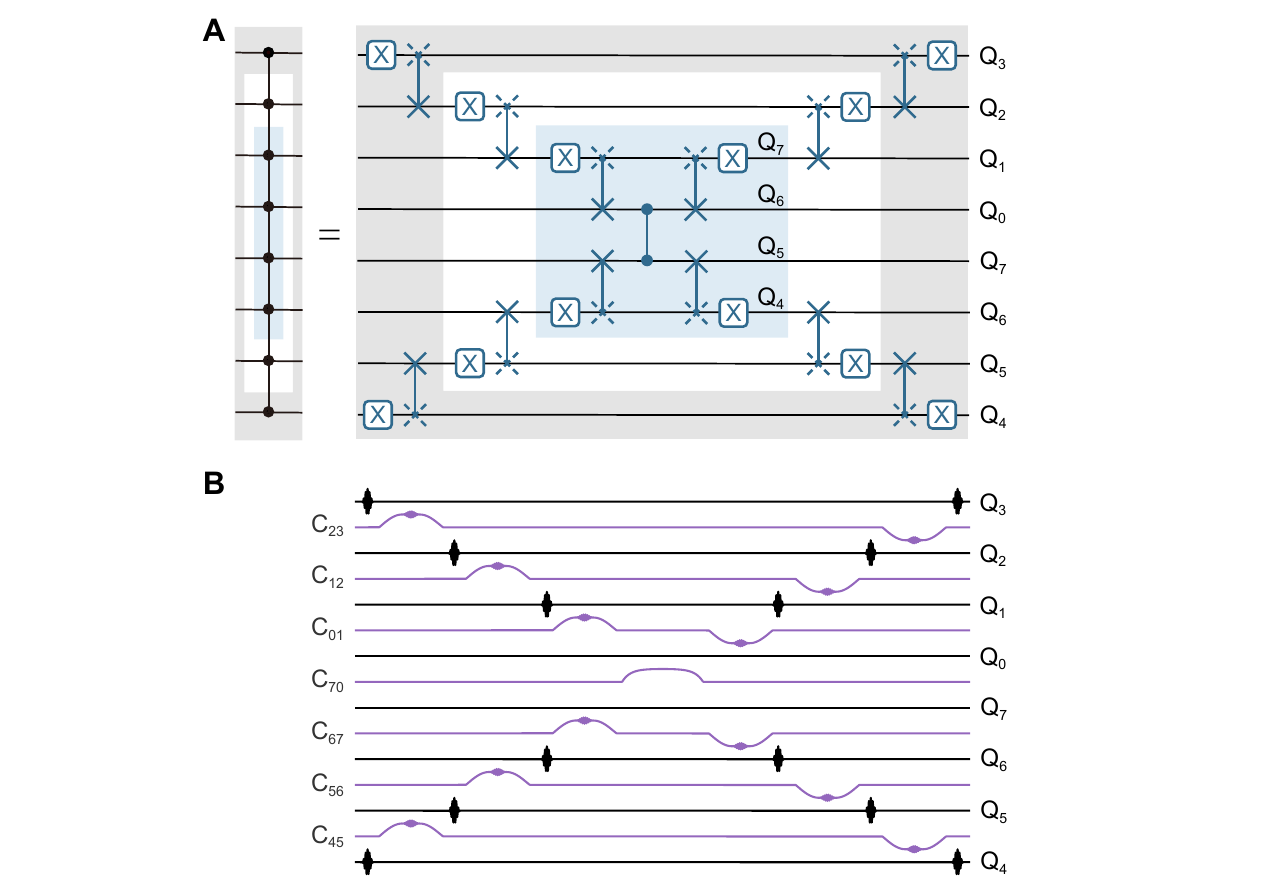}
	\caption{
		\textbf{Circuit diagram and 8-qubit pulse sequence for $n$-qubit CZ gates.} 
		\textbf{(A)} Circuit diagram for implementing the 4-qubit (blue), 6-qubit (light gray), and 8-qubit (dark gray) controlled-Z gate.
		\textbf{(B)} Experimental pulse sequence for the 8-Toffoli gate. The black (purple) curve represents qubit-XY (coupler-Z) signal. 
		The couplers are idly biased at the sweet spot and the second flux pulse is inverted for suppressing low frequency noise.
	}
	\label{fig:pulse_sequence}
\end{figure}

The $n-$CZ gate scheme can be decomposed into SWAP gates calibrated for connected qubit pairs on the 8-qubit ring and single-qubit $X$ gates.
The circuit for implementing the $n-$CZ gate with $n=4,5,8$ are illustrated in Fig.~\ref{fig:pulse_sequence}A. 
There are two considerations when selecting the qubits in each case. First, the qubit pairs at both ends of the ladder, ${\rm Q}_0$--${\rm Q}_7$ and ${\rm Q}_3$--${\rm Q}_4$, have a relatively weak qubit--coupler interaction strength ($g/2\pi \approx 40$~MHz), leading to a tighter adiabatic condition and an inferior gate performance. Second, the frequency of ${\rm Q}_3$ is unstable, showing random telegraph behavior, possibly as a result of spurious two-level systems.
The set of control waveforms for the 8-CZ gate is shown in Fig.~\ref{fig:pulse_sequence}B. 
To suppress the correlated flux noises, we idle all couplers at the sweet spot and use two-pole flux modulation for QuAND and reverse QuAND. 

\subsection{Reverse QuAND gate}

Here we discuss about the SWAP gate phase ignored in the main text and explain how we effectively calibrate it away in the experiment.
We only consider the case for SWAP between $|11\rangle$ and $|20\rangle$. The other case for SWAP between $|11\rangle$ and $|02\rangle$ can be analyzed in the same way, so not to be repeated.

\begin{figure}[htbp]
	\centering
	\includegraphics[width=170mm]{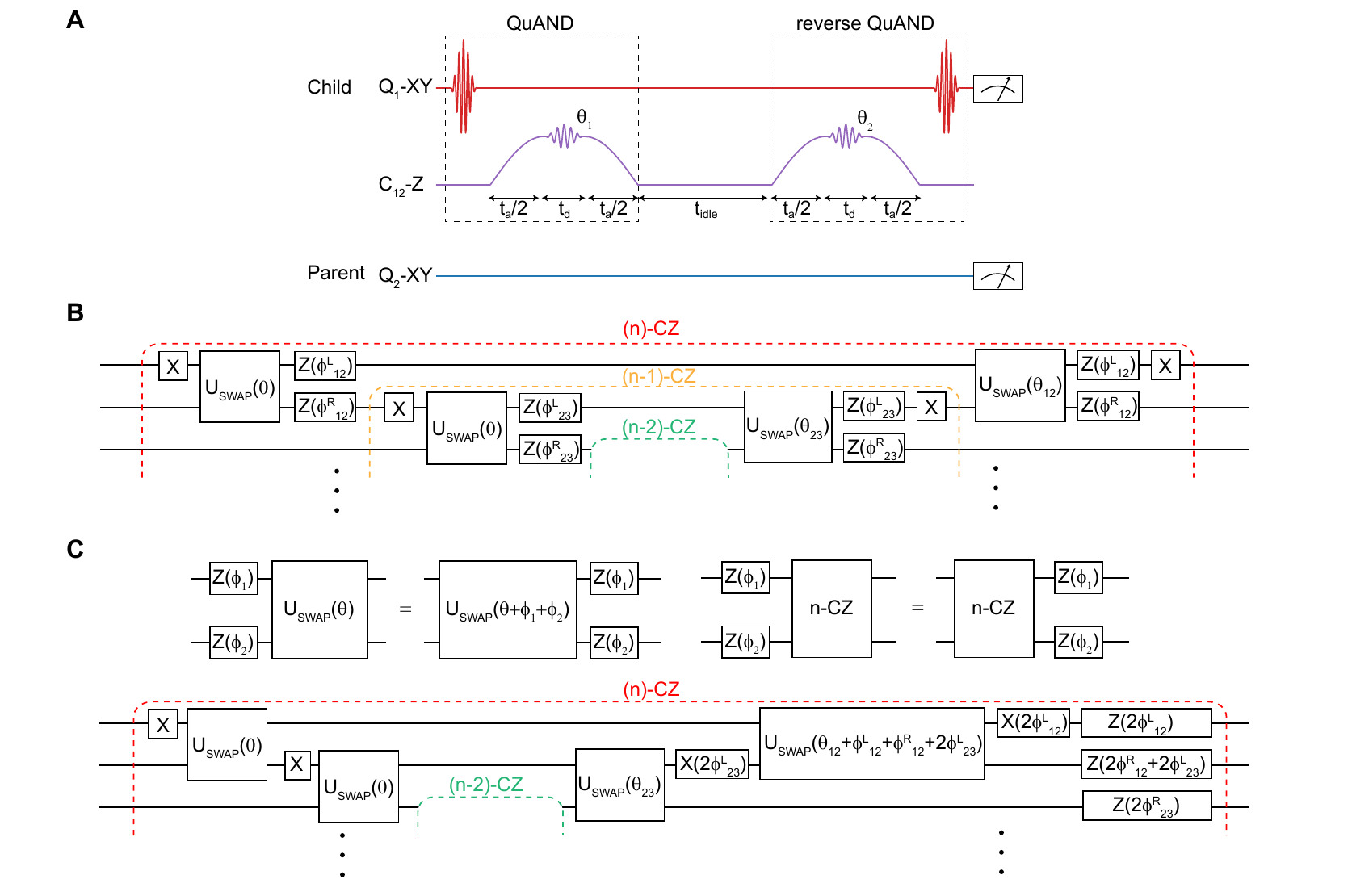}
	\caption{
		\textbf{Phase Calibration}. 
		\textbf{(A)} Pulse sequence for QuAND and reverse QuAND gate. The phase of the second parametric pulse is calibrated to cancel the conditional phase accumulation. 
		\textbf{(B)} Circuit diagram for $n$-qubit CZ gates using experimental unitary operations. The phases of the SWAP gates and the compensatory $Z$ gates are iteratively calibrated. 
		\textbf{(C)} Propagation of (virtual) $Z$ phase. We utilize the commutation relationship between $Z$ gates and SWAP or $n$-qubit CZ gates (top) to modify the circuit in (B). The phases of the $Z$ gates propagate to the end of the circuit and change the phases of subsequent $X$ rotation and SWAP operations.  
	    }
	\label{fig:phase}
\end{figure}

First, we show that the reverse ${\rm QuAND}$ can be implemented by changing the phase of the second parametric modulation.
According to the time sequence shown in Fig.~\ref{fig:phase}A, the lab-frame unitary operator for the SWAP gate between $|101\rangle$ ($|{\rm Q_1}{\rm C_{12}}{\rm Q_2} \rangle$) and $|200\rangle$ is
\begin{equation}
\begin{aligned}
	U_{\rm SWAP}(\theta) 
	=&  
	\begin{pmatrix}
		& 1 & 0 & 0 & 0 & 0 \\
		& 0 & \e^{-i(\bar{\omega}_{001}t_{\rm d} + \bar{\omega}'_{001} t_{\rm a} )} & 0 & 0 & 0 \\
		& 0 & 0 & \e^{-i(\bar{\omega}_{100}t_{\rm d} + \bar{\omega}'_{100} t_{\rm a} )} & 0 & 0 \\
		& 0 & 0 & 0 &  0 & -\i \e^{-\i [ \bar{\omega}_{101}  t_{\rm d} + (\bar{\omega}'_{101}+\bar{\omega}'_{200} )t_{\rm a}/2 + \theta] }  \\
		& 0 & 0 & 0 &  -\i \e^{-\i [\bar{\omega}_{200}  t_{\rm d} + (\bar{\omega}'_{101}+\bar{\omega}'_{200} )t_{\rm a}/2 -\theta] } & 0 \\
	\end{pmatrix},
\end{aligned}
\end{equation}
Here $t_{\rm a}$ the denotes rising and falling time of the adiabatic pulse and $t_{\rm d}$ the length of the parametric pulse.
$\omega_{s},\bar{\omega}_{s},\bar{\omega}'_{s}$ denotes the idle frequency, average frequency during the parametric modulation, average frequency during the adiabatic rising/falling edge of eigenstate $|s\rangle$.

The combined unitary (in lab frame) for two SWAP operation with an idling time $t_{\rm idle}$ in between is
\begin{equation}
\begin{aligned}
U &=U_{\rm SWAP}(\theta_2)	U_{\rm idle}   U_{\rm SWAP}(\theta_1) \\
&=\begin{pmatrix}
1 & 0 & 0 & 0 & 0 \\
0 & \e^{-i(2\bar{\omega}_{001}t_{\rm d} + \omega_{001} t_{\rm idle} + 2\bar{\omega}'_{001} t_{\rm a} )} & 0 & 0 & 0 \\
0 & 0 &  \e^{-i(2\bar{\omega}_{100}t_{\rm d} + \omega_{100} t_{\rm idle} + 2\bar{\omega}'_{100} t_{\rm a} )} & 0 & 0 \\
0 & 0 & 0 & \e^{-i (\theta_1 - \theta_2 - \pi + (\bar{\omega}'_{200} +  \bar{\omega}'_{101} )t_{\rm a} + (\bar{\omega}_{200} +  \bar{\omega}_{101} )t_{\rm d} +  {\omega}_{200} t_{\rm idle}  ) } & 0 \\
0 & 0 & 0 & 0 & * \\,
\end{pmatrix}
\end{aligned}
\end{equation}
where the asterisk indicates the phase factor on the non-computational level which we do not care about.
In the rotating frame which is equivalent to logical basis, the unitary is
\begin{equation}
\begin{aligned}
U' &= \Lambda(t_{\rm idle} + 2 t_{\rm a} + 2 t_{\rm d})^{\dagger} U  \Lambda(0) = \begin{pmatrix}
1 & 0 & 0 & 0 & 0 \\
0 & \e^{-i\phi_2} & 0 & 0 & 0 \\
0 & 0 &  \e^{-i2\phi_1} & 0 & 0 \\
0 & 0 & 0 & \e^{-i(2\phi_1+2\phi_2+\phi_{\rm zz})} & 0 \\
0 & 0 & 0 & 0 & * \\,
\end{pmatrix}, \\
{\rm where} \quad & \phi_1 = (\omega_{001} - \bar{\omega}_{001}) t_{\rm d} + ( \omega_{001} -\bar{\omega}'_{001}) t_{\rm a} ,\\
&  \phi_2 = (\omega_{100} - \bar{\omega}_{100}) t_{\rm d} + ( \omega_{100} -\bar{\omega}'_{100}) t_{\rm a}, \\
& \phi_{\rm zz} = \theta_2 -\theta_1 + \pi + (\omega_{101} - {\omega}_{200}) t_{\rm idle} + (\omega_{101}-\bar{\omega}'_{200} - \bar{\omega}'_{101} )t_{\rm a} + (\omega_{101}-\bar{\omega}_{200} -  \bar{\omega}_{101} )t_{\rm d} - (\phi_1 + \phi_2).
\end{aligned}
\end{equation}
$\phi_{\rm zz}$ is the conditional phase accumulated on state $|101\rangle$. This conditional phase comes from the adiabatic modulation and the idling part in between, so it is important to take both into account when calibrating the phase.
The phase of the first parametric pulse can be simply set to 0.
Then the reverse QuAND gate is implemented by calibrating the phase of the second parametric pulse until $\phi_{\rm zz} = 0$. The single-qubit (local) phase can be conveniently compensated by virtual \rm{Z} gates.

\subsection{Phase calibration}

The entire calibration process starts with the CZ gate in the middle, and then progressively extends to the outermost part of the circuit, as shown in Fig.~\ref{fig:phase}B.
In each iteration, we calibrate the phase of the parametric pulse in reverse QuAND through a conditional Ramsey experiment and $Z$ gate phases through single-qubit Ramsey experiments. 

We use virtual-$Z$ gates to compensate the local phases induced by the flux pulse.
Note that the SWAP gate and $Z$ gate do not commute. When swapping the order of the two operations, the phase of the $Z$ gate is absorbed into the phase of the parametric pulse.
How virtual-$Z$ phase propagate through the circuit is illustrated in Fig.~\ref{fig:phase}C.

\begin{figure}[htbp]
	\centering
	\includegraphics[width=150mm]{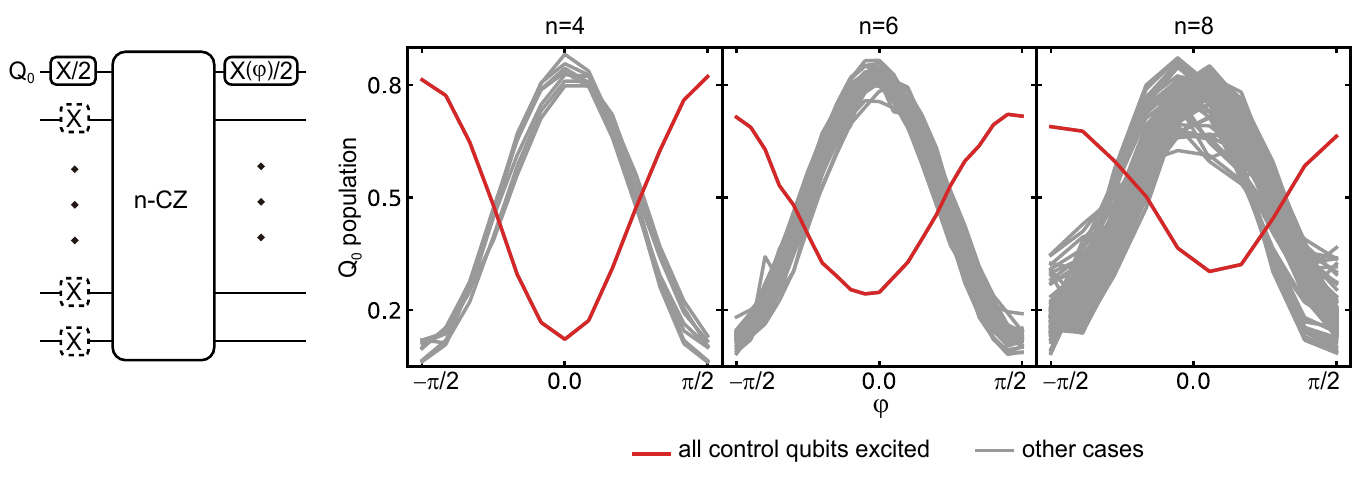}
	\caption{
		 \textbf{Multi-qubit conditional Ramsey experiments for phase checking of $n$-qubit CZ gates.} In the experiments, ${\rm Q}_0$ (target qubit) is initialized at the superposition state, while the others (control qubits) are either excited or in the ground state. The phase of the target qubit is flipped after $n$-qubit CZ if all control qubits are excited, shown as the red lines.
	}
	\label{fig:nCZ_phase}
\end{figure}

\begin{figure}[htbp]
	\centering
	\includegraphics[width=120mm]{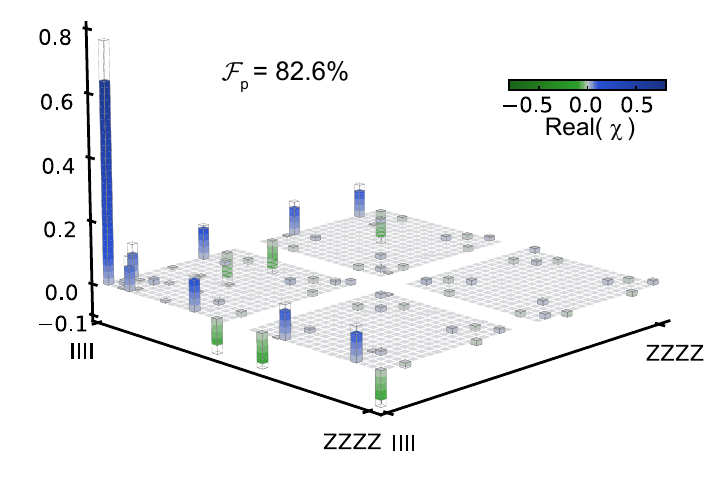}
	\caption{
		\textbf{Process tomography of the 4-CZ or CCCZ gate (color bar: experiment; frame: theory).}
	}
	\label{fig:process_tomo}
\end{figure}

The phase calibration is verified by multi-qubit conditional Ramsey experiments (Fig.~\ref{fig:nCZ_phase}). 
The circuit diagram is shown on the left. $Q_0$ is chosen as the target qubit in all the shown cases ($n=4,6,8$) while the other qubits work as the control ones.
There is approximately a $\pi$ phase shift on the target qubit when all the control qubits are in the excited state, shown as the red lines, verifying the phase calibration.

To characterize the synthesized gate, we perform standard process tomography for the 4-qubit CZ or CCCZ gate (larger matrix is unattainable due to limited memory). 
The reconstructed process matrix $\chi_{\rm exp}$ from $4^4=256$ distinct input states gives process fidelity $\mathcal{F}_{\rm p} = \rm{Tr}(\chi_{\rm exp} \chi_{\rm ideal}) = $82.6\% (Fig.~\ref{fig:process_tomo}).
Note that the process fidelity of the 4-qubit CZ is lower then the 4-qubit Toffoli truth-table fidelity shown in the main text (both using $\rm{Q}_4\rm{Q}_5\rm{Q}_6\rm{Q}_7$), partly as a result of underestimated phase errors in the truth-table measurements.


\section{Toffoli gate error}

\begin{figure}[htbp]
	\centering
	\includegraphics[width=100mm]{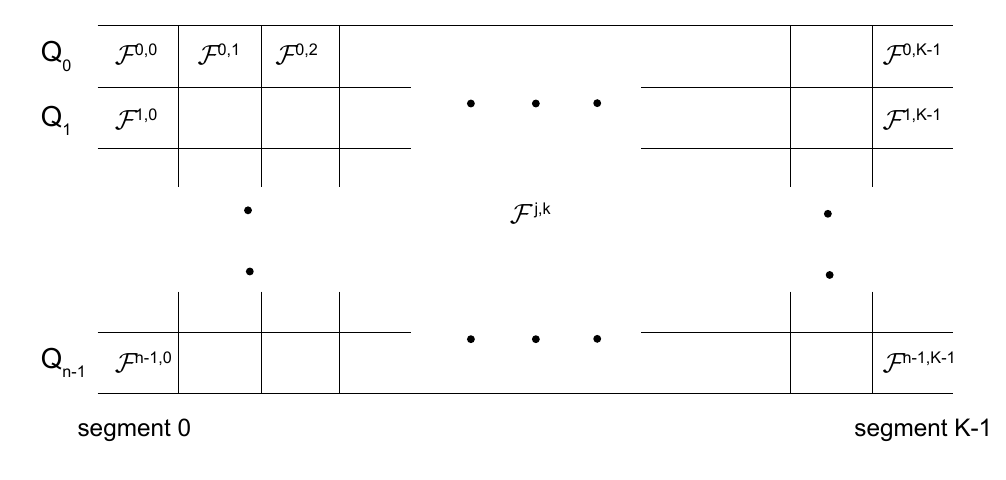}
	\caption{
		\textbf{Fidelity analysis and pulse sequence for $n$-qubit Toffoli.}
	    Sketch of segmented circuit for fidelity estimate.
		$\mathcal{F}_{\rm T1}^{j,k}$ represents T1-limited fidelity of qubit $j$ in segment $k$.
		}
	\label{fig:T1_error}
\end{figure}

We estimate the T1-limited gate fidelity ($\mathcal{F_{\rm T1}}$) by segmenting the circuit (in Fig.~\ref{fig:T1_error}) and taking the product of T1-limited fidelity of each segment $\mathcal{F_{\rm T1}} =\Pi_{j,k}  \mathcal{F}_{\rm T1}^{j,k}$.
The relaxation rate of an instantaneous eigenstate during a flux pulse generally keeps varying during the pulse, due to the varying wavefunction participation of different bare states \cite{chu2021coupler}. The average relaxation rate of the eigenstate is calculated by summing up contribution from all bare states. The rate is larger than that of idle status, because of the strong overlap with the relatively short-lived coupler (see Table.~\ref{table:Device}).

The relaxation-limited gate fidelities (total duration) for the 4-qubit, 6-qubit, and 8-qubit Toffoli gates are 92.5\% (0.4~$\mu$s), 66.7\% (1.3~$\mu$s, staggered pulses), and 62.3\% (1.1~$\mu$s), respectively, and are responsible for approximately 70\% of the total error.
Note that the gate length of 6-Toffoli is longer because we stagger the flux pulses for reducing the otherwise strong spectator effect \cite{chu2021coupler}.

\section{Grover's Search Algorithm}	

The Grover's search algorithm generally includes four steps (as shown in Fig.~3A in the main text): 
(\romannumeral 1) initialize the $n$-qubit system into an equal superposition of all possible bit-string states $\frac{1}{2^{n/2}}\sum_{s=0}^{2^n-1} \ket{s}$;
(\romannumeral 2) encode the solution $j$ with a phase oracle $O_{j} = \sum_{s \neq j} \ket{s}\bra{s} - \ket{j}\bra{j}$, i.e.\ a conditional $\pi$-phase shift on state \ket{j};
(\romannumeral 3) diffuse the encoded phase and amplify the probability of finding \ket{j}; 
(\romannumeral 4) measure the final state.
Step (\romannumeral 2) and (\romannumeral 3) may be repeated for $M$ times for further amplification. The algorithm promises quadratic speedup, reaching the optimal amplification at $M=2^{n/2}$.

\subsection{Error model for algorithm success probability}
Here we provide a simple model for estimating the success probability (ASP) in Grover's search algorithm with non-ideal gate fidelity.
To find one solution among an unstructured list of size $2^n$ using Grover's algorithm, the ideal ASP after $M$ cycles of oracle queries is $\sin^2\left((2M+1)\arcsin(2^{-\frac{n}{2}})\right)$. When the coherence is completed destroyed, the measurement outcome is a uniformly random number between $0$ and $1$ because the last layer of the circuit is a layer of Hadamard gates. Hence, the ASP is $\frac{1}{2^n}$. Since the circuit of each encoding-diffusing cycle contains two $n$-qubit CZ gates with fidelity $\mathcal{F}$ (ignoring single-qubit gate error), the fidelity of the whole circuit with $M$ cycles is then $\mathcal{F}^{2M}$. A empirical ASP model can be written 
\begin{equation}
\rm{ASP}=\mathcal{F}^{2M}\sin^2\left((2M+1)\arcsin(2^{-\frac{n}{2}})\right)+\frac{1-\mathcal{F}^{2M}}{2^n}.
\end{equation}

\subsection{Multi-solution Grover's Search}

\begin{figure}[htbp]
	\centering
	\includegraphics[width=170mm]{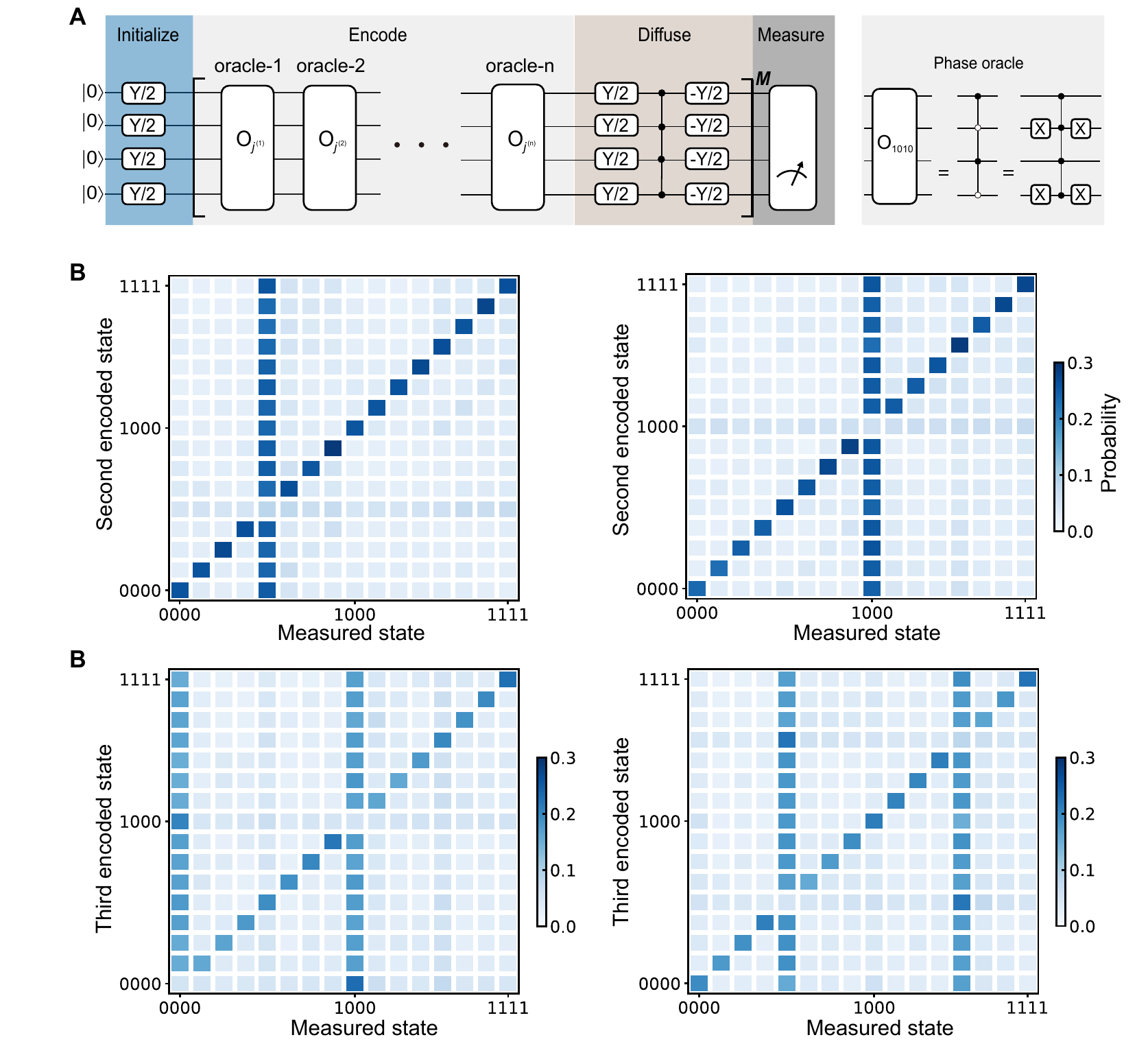}
	\caption{
		\textbf{Multiple solution Grover's search algorithm}. \textbf{A} Circuit diagram for implementing  $n$-solution Grover's search algorithm. During the encoding process, we apply $n$ phase oracles in succession. An example of a 4-qubit phase oracle is shown on the right. \textbf{B} Four-qubit two-solution Grover's search result ($M=1$). 
		With the first encoded state fixed as $|0100\rangle$ (left) or $|1000\rangle$ (right), the matrix shows how the measured probabilities of $2^4=16$ states vary with the second encode state (y-axis).
		When state $|0000\rangle$ is encoded twice (bottom), the net effect equals to no encoding and all states are measured with about equal probability. 
		\textbf{C} Four-qubit three-solution Grover's search result ($M=1$). The first, second encoded states are fixed as $|0000\rangle$,$|1000\rangle$ (left) or $|0100\rangle$,$|1100\rangle$ (right) . The third one traverses over all the 16 logical states (y-axis).
	}
	\label{fig:mgrover}
\end{figure}

Arbitrary multiple solutions can be conveniently encoded with concatenated oracles, as shown in Fig.~\ref{fig:mgrover}A.
Examples of experimental results for two-solution and three-solution Grover's search are shown in Fig.~\ref{fig:mgrover}B and C.

\bibliographystyle{apsrev4-1}
\bibliography{BibDoi}